\documentclass[aps,prl,reprint,superscriptaddress,nofootinbib,nobibnotes,longbibliography]{revtex4-2}
\usepackage{amsmath,amssymb,bm,mathtools,graphicx,microtype}
\usepackage[hidelinks]{hyperref}
\hypersetup{pdftitle={Reservoir Support Sets the Minimum Loss for Non-Hermitian Dynamics},pdfauthor={Alena Mastiukova},pdfsubject={Phase-controlled passive reservoir synthesis and support certification}}
\newcommand{\Kbar}{\overline K}
\newcommand{\one}{\mathbb I}
\newcommand{\Tr}{\operatorname{Tr}}
\newcommand{\Iop}{\operatorname{Im}_{\mathrm{op}}}
\makeatletter
\AtBeginDocument{\def\NAT@cmprs{\z@}}
\makeatother
\begin{document}
\title{Reservoir Support Sets the Minimum Loss for Non-Hermitian Dynamics}
\author{Alena Mastiukova}
\email[Contact author: ]{a.mastiukova@rqc.ru}
\affiliation{National University of Science and Technology ``MISIS'', Moscow 119049, Russia}
\affiliation{Russian Quantum Center, Skolkovo, Moscow 121205, Russia}
\date{September 9, 2026}
\begin{abstract}
The minimum loss required to reproduce non-Hermitian dynamics depends on how many physical modes each reservoir can address coherently. Established positive-matrix criteria bound this cost. We construct a phase-controlled cubic-root Su-Schrieffer-Heeger chain in which three-mode reservoirs attain the unrestricted passivity threshold throughout its phase diagram and halve the minimum pair-supported loss at a tuned point. The bound constrains exact conditional trajectories, including time-dependent Markov controls, rather than preparation of one final state. A connected-chain protocol tests effective reservoir support through calibrated finite-time emission. Full system-auxiliary propagation then bounds conditional fidelity and target yield for every one-particle input. A coherent controller that accurately prepares the selected endpoint fails this propagator test. A singular-value bound excludes every unitary realization of the same conditional map. These results connect local reservoir access to a measurable loss cost for prescribed dynamics.
\end{abstract}
\maketitle

Non-Hermitian lattices describe directional transport, root spectra, and exceptional dynamics~\cite{Kawabata2019,Bergholtz2021}. To realize such evolution passively, the no-jump generator must be compatible with completely positive Lindblad dynamics. The resulting minimum uniform loss is fixed by the numerical abscissa of the evolution generator, rather than only its eigenvalues~\cite{McDonald2022,Okuma2022}. Experiments impose a second constraint: one reservoir may coherently address only a few physical modes. What loss penalty does this restricted access impose, and can a measured signal certify that the restriction has been overcome?

We consider static, Markovian effective generators with linear annihilation losses in a fixed orbital basis. Factor-width and comparison-matrix criteria~\cite{Boman2005,Johnston2025}, together with their multilevel-coherence formulation~\cite{Ringbauer2018,Kraft2021}, provide a sharp bound on the reduction in loss attainable with three-mode reservoirs. A phase-controlled cubic-root SSH chain~\cite{SSH1979,Viedma2024,McCann2026} reaches this bound with explicit local channels. We connect this lattice construction to calibrated emission measurements and test the entire one-particle propagator in a finite-bandwidth reservoir model. This distinction matters: preparing one endpoint can be much easier than reproducing the prescribed dynamics.

\paragraph{A sharp cost of restricted support.}
For $L_\mu=\sum_aB_{\mu a}c_a$, the one-particle no-jump matrix is $K=H-iM/2$, with $H=H^\dagger$ and $M=B^\dagger B\succeq0$. A prescribed target fixes
\begin{equation}
 K_\Gamma=\Kbar-i\Gamma\one/2,\quad
 M_\Gamma=\Gamma\one+F,\quad F=i(\Kbar-\Kbar^\dagger).
 \label{eq:target}
\end{equation}
The operator imaginary part is
\begin{equation}
 A\equiv\Iop(\Kbar)=\frac{\Kbar-\Kbar^\dagger}{2i}=A^\dagger,
 \qquad F=-2A.
 \label{eq:operatorimag}
\end{equation}
Here $\dagger$ denotes conjugate transpose. The symbol $\Iop$ is not the entrywise imaginary part. The unrestricted threshold is $\Gamma_{\rm CP}=-\lambda_{\min}(F)=2\lambda_{\max}(A)$. Here $\lambda_{\max}(A)$ is the numerical abscissa of $-i\Kbar$~\cite{McDonald2022,Okuma2022}.

Hereafter assume $F_{aa}=0$, so every diagonal element of $M_\Gamma$ equals $\Gamma$. Let $\Gamma_q$ be the minimum shift with at most $q\geq2$ nonzero scalar-mode coefficients in each row of $B$. The number of rows, their phases and amplitudes, and onsite balancing losses are unrestricted. For $W_{ab}=|F_{ab}|$ off the diagonal and $W_{aa}=0$, the comparison-matrix criterion gives $\Gamma_2=\rho(W)$~\cite{Boman2005,Ringbauer2018}. Moreover,
\begin{equation}
 \boxed{\Gamma_q\leq\Gamma_2\leq(q-1)\Gamma_q.}
 \label{eq:sharp}
\end{equation}
This is the reservoir-cost form of the established conversion of higher-support positive matrices into pair-supported ones~\cite{Ringbauer2018}. The bound follows directly from the support constraint. For every real $x\geq0$ and admissible factor,
\begin{align}
 x^TWx&\leq\sum_\mu\left[
 (\sum_a|B_{\mu a}|x_a)^2-\sum_a|B_{\mu a}|^2x_a^2\right]\nonumber\\
 &\leq(q-1)\Gamma\|x\|^2.
 \label{eq:sharp-proof}
\end{align}
The last step uses the row support and $M_{aa}=\Gamma$. Choosing a Perron vector and minimizing $\Gamma$ proves the upper bound. Inclusion of the support classes proves the lower bound. Thus three-mode channels can reduce the optimal pair-supported loss by at most 50\% for a prescribed generator and a uniform shift. Constructive proofs, nonuniform diagonals, and sharp examples are given in the Supplemental Material (SM)~\cite{Supplemental}.

The same cost applies to exact conditional trajectories under time-local Markov controls. If a differentiable propagator $V$, with $V(0)=\one$, gives the same normalized output as $T(s)=e^{-i\Kbar s}$ for every pure input and every $0\leq s\leq\tau$, linearity forces $V=a(\tau)T$. Its generator differs from $\Kbar$ only by a scalar, so instantaneous support at most $q$ requires $-2\partial_\tau\log|a|\geq\Gamma_q$ and $|a(\tau)|^2\leq e^{-\Gamma_q\tau}$. This statement does not follow from agreement on one input or at one endpoint (SM).

\begin{figure*}[t]
 \includegraphics[width=\textwidth]{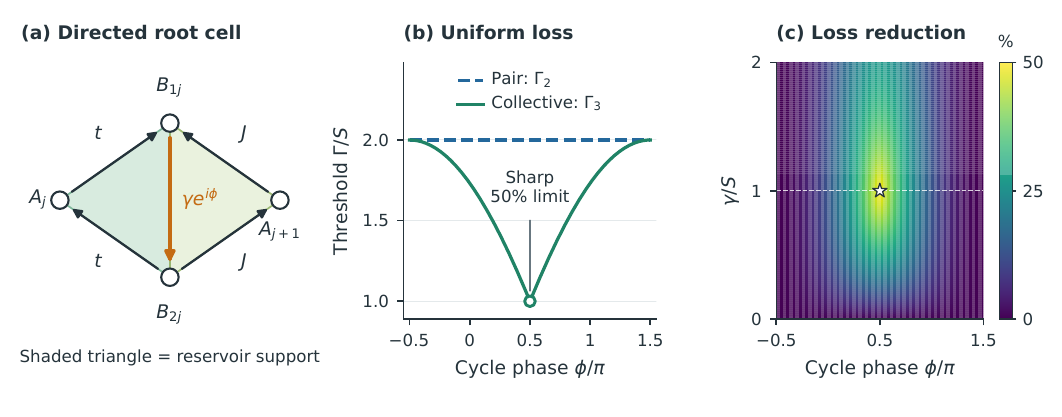}
 \caption{\label{fig:phase}Saturating the support-cost bound. (a) Directed cubic-root cell and the two reservoir-support triangles. (b) Exact thresholds at $\gamma=S=t+J$. The marked point saturates the 50\% bound on loss reduction when two-mode reservoirs are replaced by three-mode reservoirs. (c) Exact reduction $1-\Gamma_3/\Gamma_2$ versus phase and $\gamma/S$. The architectures realize the same target at each point. Changing the directed-cycle phase changes that target.}
\end{figure*}

\begin{figure*}[t]
 \includegraphics[width=\textwidth]{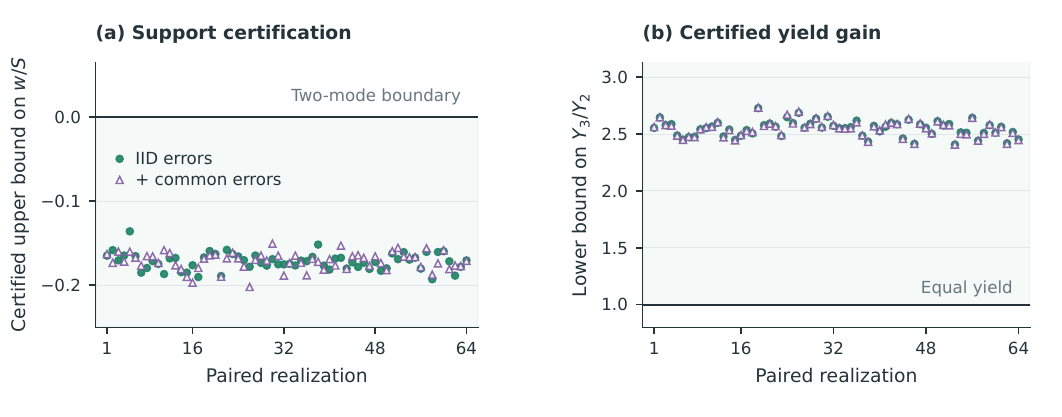}
 \caption{\label{fig:witness}Support certification and target yield on a connected chain. For each realization at $L=6$ and $\kappa/S=1920$, the same collective-reservoir device supplies both datasets. (a) Corrected upper witness bounds. Negativity excludes every effective pair-supported loss matrix in the fixed basis. (b) Lower bounds on $Y_3/Y_2$ from target-projection counts for both architectures. Each series contains 64 draws. The correlated series augments the same independent-error draws with a common relative phase and auxiliary detuning. All counts are synthetic. The bounds have 99\% simultaneous coverage over all 5120 binomial groups, conditional on the stated calibration bounds.}
\end{figure*}

\paragraph{Local saturation in a root lattice.}
Introduce a cycle phase into the cubic-root chain,
\begin{equation}
 \Kbar_\phi(k)=\begin{pmatrix}0&0&h^*\\h&0&0\\0&\gamma e^{i\phi}&0\end{pmatrix},
 \quad h=t+Je^{ik},\quad S=t+J,
 \label{eq:phase}
\end{equation}
where $t,J,\gamma>0$. Its cube is $e^{i\phi}\gamma|h|^2\one$. The identity $\Kbar_\phi=e^{i\phi/3}D_\phi\Kbar_0D_\phi^\dagger$, with $D_\phi=\operatorname{diag}(1,e^{-i\phi/3},e^{i\phi/3})$, preserves the root eigenvectors up to onsite rephasing and leaves Jordan-block sizes unchanged. The overall complex rotation changes passivity relative to the fixed decay half-plane: the cycle phase cannot be removed by an orbital gauge transformation.

Pair-supported synthesis discards the edge phases in $W$, giving
\begin{equation}
 \Gamma_2=\frac{\gamma+\sqrt{\gamma^2+8S^2}}2.
 \label{eq:pairbulk}
\end{equation}
By contrast, the eigenvalues of $F(k)$ obey
\begin{equation}
 \lambda^3-(2|h|^2+\gamma^2)\lambda-2\gamma|h|^2\sin\phi=0.
 \label{eq:cubic}
\end{equation}
The lowest eigenvalue occurs at $|h|=S$. Define $C=\Gamma_{\rm CP}\one+F(h=S)\succeq0$. Embed $(t/S)C$ on $(A_j,B_{1j},B_{2j})$ and $(J/S)C$ on $(A_{j+1},B_{1j},B_{2j})$. Their sum is exactly $M_{\Gamma_{\rm CP}}$. Factoring these triangles establishes $\Gamma_3=\Gamma_{\rm CP}$ for every phase and $\gamma/S$. At most four periodically repeated channel families suffice.

At $\phi=\pi/2$, $\gamma=S$, each triangle factor has rank one:
\begin{align}
 L_{tj}&=\sqrt t\,(c_{Aj}-ic_{B_1j}+ic_{B_2j}),\nonumber\\
 L_{Jj}&=\sqrt J\,(c_{A,j+1}-ic_{B_1j}+ic_{B_2j}).
 \label{eq:triangles}
\end{align}
These two families give $\Gamma_3=S$ and $\Gamma_2=2S$, saturating Eq.~\eqref{eq:sharp}. The loss matrix has rank two at generic momentum, so two families are also necessary for a translation-invariant realization. This bulk count differs from the number of auxiliaries in a finite chain. The real-phase baseline permits only a 13.40\% maximum reduction, while cycle-phase control reaches the general 50\% ceiling [Fig.~\ref{fig:phase}].

The periodic tuned model has undamped modes at $k=0$. Adding common onsite loss $\delta>0$ retains a saving $S/(2S+\delta)$, or 47.62\% at $\delta=0.1S$. For an open chain let $s=\|t\one+J\mathsf S_L\|_2<S$. At $\gamma=S$, its exact thresholds are $\Gamma_3^{(L)}=S$ and $\Gamma_2^{(L)}=(S+\sqrt{S^2+8s^2})/2$. The case $t/J=0.5$, $L=6$ already gives 49.08\% with a positive finite-size relaxation gap. The support penalty depends on $F$ and can occur for normal targets as well: it is distinct from the numerical-abscissa excess associated with nonnormality.

\begin{figure*}[t]
 \includegraphics[width=\textwidth]{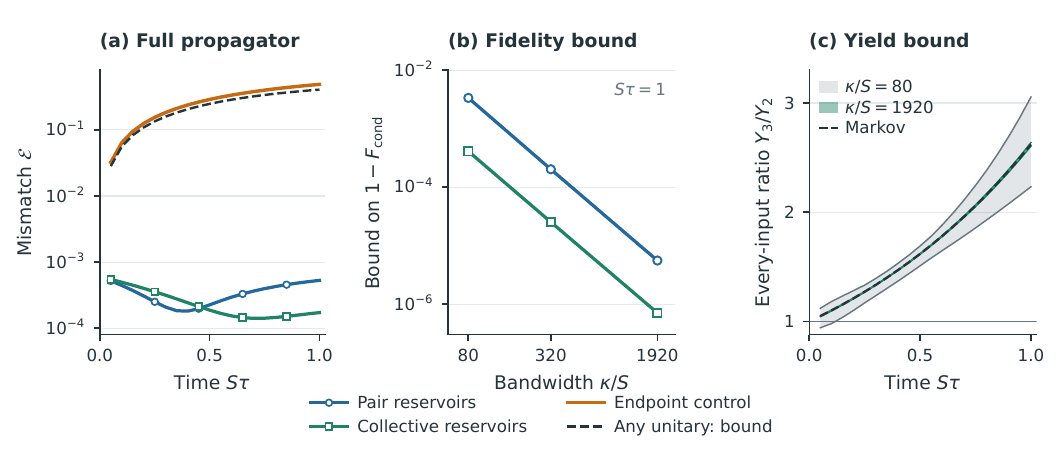}
 \caption{\label{fig:matched}Testing prescribed dynamics on the nominal six-cell device. (a) Propagator mismatch $\mathcal E$ at $\kappa/S=1920$, including the endpoint-optimized coherent controller and the lower bound for every unitary. (b) Upper bounds $r_r^2$ on conditional infidelity for every input at $S\tau=1$ versus bandwidth. The coupling capability changes with bandwidth as $G\propto\sqrt\kappa$. (c) Input-uniform enclosures of $Y_3/Y_2$ at two bandwidths. The dashed curve is the exact Markov ratio. Bands are matrix-norm bounds evaluated in the stated model, not statistical intervals. Curves connect the computed time points.}
\end{figure*}

\paragraph{Certifying the support on the same device.}
For a normalized Perron vector $v$ of $W$, define $Q_{aa}=v_a^2$ and $Q_{ab}=-e^{i\arg F_{ab}}v_av_b$ on each edge, with zero otherwise. Every two-dimensional principal block of $Q$ is positive semidefinite, hence
\begin{align}
 w(B^\dagger B)&\equiv\Tr(QB^\dagger B)\geq0
 \quad\text{for every pair-supported }B,\nonumber\\
 w(M_\Gamma)&=\Gamma-\Gamma_2.
 \label{eq:witness}
\end{align}
This construction is a dual-cone witness~\cite{Ringbauer2018}. In the Markov limit, one-particle loss rates $r(\psi)=\langle\psi|M|\psi\rangle$ reconstruct $w$. Localized and pair-superposition inputs need $N+E$ settings. Alternatively, the spectral decomposition $Q=\sum_\nu q_\nu|u_\nu\rangle\langle u_\nu|$ needs $N$ generally extended inputs.

The finite-bandwidth implementation couples empty auxiliary modes through $G=\sqrt\kappa B/2$ and damps them at linewidth $\kappa$~\cite{Metelmann2015,LiuSegal2020}:
\begin{equation}
 K_{\rm ext}=\begin{pmatrix}H&G^\dagger\\G&D_{\rm aux}-i\kappa\one/2\end{pmatrix},
 \qquad D_{\rm aux}=D_{\rm aux}^\dagger.
 \label{eq:aux}
\end{equation}
For initially empty auxiliaries, the instantaneous emission rate vanishes and cannot be identified with $r(\psi)$. We instead use emitted-particle probabilities at two finite times, with a uniform correction bound for reservoir memory, detuning, and state evolution. Exact binomial intervals account for counting uncertainty, and calibration bounds account for preparation and detection errors. A corrected upper bound $w_{\rm up}<0$ certifies that the effective Markov loss matrix requires support beyond pairs in the fixed basis, conditional on the calibrated auxiliary model.

For the connected $L=6$ chain at fixed $\kappa/S=1920$, each collective-device realization supplies both the certificate and the target-yield data in Fig.~\ref{fig:witness}. The protocol uses 18 generally extended witness inputs and $3.6\times10^7$ synthetic trials per device. Every corrected upper bound is negative in both the 64 independent-error draws and their correlated extension with a common phase error and auxiliary detuning. The 99\% simultaneous coverage concerns counting noise and is conditional on the supplied calibration bounds.
The two architectures share the nominal target, coherent couplings, linewidth, available auxiliary bank, and coupling ceiling. The collective construction activates 13 auxiliaries. The optimal pair factor activates 28 in the same 28-slot bank. The SM reports coupling power and required control precision separately. The specified hardware ceilings remain fixed throughout this test.

\paragraph{Readout, robustness, and the scope of optimization.}
Starting from a central $A$ orbital, define the normalized target state $|\psi_{\rm tar}(\tau)\rangle\propto e^{-i\Kbar\tau}|\psi_0\rangle$. We evaluate the target-projection probability $Y=|\langle\psi_{\rm tar}|u_{\rm sys}\rangle|^2$, system survival $P_{\rm sys}=\|u_{\rm sys}\|^2$, and conditional fidelity $F_{\rm cond}=Y/P_{\rm sys}$. An explicit sequence of two-mode rotations maps the target state to one output orbital. Detection there implements the projection. Applying the same construction separately to each witness eigenvector and reversing that sequence prepares the corresponding input. The rotations assume switchable reservoir coupling during preparation and readout. Residual gate errors enter the stated calibration allowances. The device test uses numerical propagation and synthetic counts.

For the exact Markov target family, every input has the one-particle yield ratio $Y_3/Y_2=e^{(\Gamma_2-\Gamma_3)\tau}$, with the attainable shifts fixed by the support constraints. Full propagation of Eq.~\eqref{eq:aux} quantifies finite-bandwidth and error corrections to this relation. For the noisy device ensemble at $S\tau=1$ the independent-error ensemble gives a median $Y_3/Y_2=2.625$ with empirical 5-95\% quantiles $[2.501,2.726]$. The correlated extension gives comparable results. Count-based lower bounds on the yield ratio exceed $2.40$ in all 128 cases. Both conditional fidelities, computed from the propagated states, exceed $0.9985$. These count intervals do not independently certify the stated fidelities.
If target projection defines success and preparation, readout, and cycle durations are equal, the yield ratio is also the reduction in expected trials per success. The numerical comparison applies to the specified finite ensemble and calibration bounds.

\paragraph{Prescribed dynamics beyond one endpoint.}
The loss advantage concerns a map, not an isolated output state. Indeed, independently adjusting only the existing coherent-link amplitudes, with the same ceiling $\max|H_{ab}|/S\leq0.55$ and no reservoirs, gives $Y=F_{\rm cond}=0.999025$ for the selected endpoint (SM). We therefore compare the system propagator $V_r$ from Eq.~\eqref{eq:aux} with $T=e^{-i\Kbar\tau}$ using
\begin{equation}
 a_r=\frac{\Tr(T^\dagger V_r)}{\|T\|_F^2},\qquad
 \mathcal E_r=\frac{\|V_r-a_rT\|_F}{\|V_r\|_F}.
 \label{eq:maperror}
\end{equation}
This scalar-invariant residual tests all matrix elements. For every unitary $U$ on the system modes, the singular-value trace inequality gives
\begin{equation}
 \mathcal E_U\geq\sqrt{1-\frac{\|T\|_*^2}{N\|T\|_F^2}},
 \label{eq:unitaryfloor}
\end{equation}
where $\|T\|_*$ is the sum of singular values. The polar unitary saturates the bound~\cite{Higham1986}. At $S\tau=1$ this floor is $0.4069$, while the endpoint controller has $\mathcal E=0.4885$. The nominal pair and collective devices at $\kappa/S=1920$ give $5.29\times10^{-4}$ and $1.72\times10^{-4}$, respectively [Fig.~\ref{fig:matched}(a)].

Matrix norms also control every input, without identifying a sampled fidelity minimum with a worst-case result. For each input $\psi$, the target is $\psi_{\rm tar}(\psi)=T\psi/\|T\psi\|$ and $Y_r(\psi)=|\langle\psi_{\rm tar}(\psi)|V_r\psi\rangle|^2$. Set $r_r=\|V_r/a_r-T\|_2/\sigma_{\min}(T)$. For $r_2,r_3<1$,
\begin{align}
 F_{{\rm cond},r}(\psi)&\geq1-r_r^2,\nonumber\\
 \frac{Y_3(\psi)}{Y_2(\psi)}&\geq
 \left|\frac{a_3}{a_2}\right|^2
 \left(\frac{1-r_3}{1+r_2}\right)^2.
 \label{eq:allinput}
\end{align}
At $S\tau=1$, $\kappa/S=1920$, these give $F_{{\rm cond},2}\geq0.9999944$, $F_{{\rm cond},3}\geq0.9999992$, and $Y_3/Y_2\geq2.6043$ for every input. Even at $\kappa/S=80$, both fidelity bounds exceed $0.9966$ and the yield ratio exceeds $2.2335$. These are bounds for nominal finite-bandwidth devices, separate from the noisy count certificate in Fig.~\ref{fig:witness}. The SM provides proofs, 30 time points, and checks on 1160 input states.

The root spectrum, conditional clock encoding, and edge exceptional dynamics motivate the target. Their connections to earlier work are developed in the SM~\cite{Mastiukova2026,Wiersig2020,WuEtAl2026,GhoshBhattacharya2026,XuYi2026}. Reservoir support imposes a sharp loss constraint beyond unrestricted passivity. The root lattice saturates the three-mode bound, while calibrated emission and input-uniform propagator bounds distinguish an attainable simulation advantage from an endpoint-preparation shortcut.

\begin{acknowledgments}
The work was supported by the Ministry of Science and Higher Education of the Russian Federation in the framework of the Program of Strategic Academic Leadership ``Priority 2030'' (MISIS Strategic Technology Project Quantum Internet).

\end{acknowledgments}

\paragraph{Data availability.}
The numerical data underlying the figures and the code used for the calculations and figure generation are available from the corresponding author upon reasonable request.
\nocite{Alcaraz2021,Bai2026,Baxter1989,Boman2005,Cobanera2014,Ephremidze2009,Ephremidze2015,Fendley2014,GhoshBhattacharya2026,Hubisz2021,Johnston2025,JohnstonCoherence2022,LeRegent2024,Lee2023,Li2025,Liu2025,LiuSegal2020,Mastiukova2026,McCann2026,McCannFock2026,McDonald2022,Metelmann2015,Minganti2019,Minganti2020,Okuma2022,Ringbauer2018,Shiralieva2026,Torres2014,Traverso2023,Wanjura2025,Wiersig2020,WiersigChen2025,Xiang2026,XuYi2026}
\bibliography{references}
\end{document}


\title{Supplemental Material for ``Reservoir Support Sets the Minimum Loss for Non-Hermitian Dynamics''}
\author{Alena Mastiukova}
\email[Contact author: ]{a.mastiukova@rqc.ru}
\affiliation{National University of Science and Technology ``MISIS'', Moscow 119049, Russia}
\affiliation{Russian Quantum Center, Skolkovo, Moscow 121205, Russia}
\date{September 9, 2026}
\maketitle

\setcounter{equation}{0}
\renewcommand{\theequation}{S\arabic{equation}}
\setcounter{figure}{0}
\renewcommand{\thefigure}{S\arabic{figure}}
\setcounter{table}{0}
\renewcommand{\thetable}{S\arabic{table}}

This Supplemental Material first presents the proofs and protocols underlying the Letter, followed by related symmetry, spectral, implementation, and numerical results. Calculations for the real-phase model are marked $\phi=0$ and use the parameters stated in each section. All measurement records are synthetic, and all performance results are numerical predictions.

\medskip
\begingroup
\renewcommand{\arraystretch}{1.18}
\centering
\begin{tabularx}{\linewidth}{@{}Xr@{}}
\toprule
\textbf{Contents} & \textbf{Page} \\
\midrule
\multicolumn{2}{@{}l}{\textit{Core results and protocols}} \\
Passivity threshold and numerical range & \hyperref[S:sec:passivity]{\pageref*{S:sec:passivity}} \\
Sharp bound on reservoir-support advantage & \hyperref[S:sec:support]{\pageref*{S:sec:support}} \\
Phase-controlled support advantage & \hyperref[S:sec:phase]{\pageref*{S:sec:phase}} \\
Single-particle certification of collective loss & \hyperref[S:sec:witness]{\pageref*{S:sec:witness}} \\
Connected device: certification and target yield & \hyperref[S:sec:device]{\pageref*{S:sec:device}} \\
Acquisition and calibration budgets & \hyperref[S:sec:resource-budget]{\pageref*{S:sec:resource-budget}} \\
Matched finite-bandwidth architecture comparison & \hyperref[S:sec:bandwidth]{\pageref*{S:sec:bandwidth}} \\
Prescribed dynamics and input-uniform bounds & \hyperref[S:sec:projective-dynamics]{\pageref*{S:sec:projective-dynamics}} \\
Finite-time scalar controls & \hyperref[S:sec:controls]{\pageref*{S:sec:controls}} \\
\addlinespace[5pt]
\multicolumn{2}{@{}l}{\textit{Connections, realization, and numerical details}} \\
Origin-centered and affine complex-chiral symmetry & \hyperref[S:sec:affine]{\pageref*{S:sec:affine}} \\
Liouvillian spectrum and parity-resolved gaps & \hyperref[S:sec:liouvillian]{\pageref*{S:sec:liouvillian}} \\
Local saturation in one dimension & \hyperref[S:sec:local]{\pageref*{S:sec:local}} \\
Real-phase cubic-root SSH realization & \hyperref[S:sec:realphase]{\pageref*{S:sec:realphase}} \\
Biorthogonal fixed-occupancy clock sector & \hyperref[S:sec:clock]{\pageref*{S:sec:clock}} \\
Exceptional block in the decaying density sector & \hyperref[S:sec:exceptional]{\pageref*{S:sec:exceptional}} \\
Lossy-auxiliary implementation and robustness & \hyperref[S:sec:implementation]{\pageref*{S:sec:implementation}} \\
Numerical and plotting details & \hyperref[S:sec:numerics]{\pageref*{S:sec:numerics}} \\
\bottomrule
\end{tabularx}
\par
\endgroup
\clearpage

\section{Passivity threshold and numerical range}
\label{S:sec:passivity}

\subsection{Finite-dimensional theorem}

Consider a finite set of fermionic orbitals with a number-conserving quadratic Hamiltonian
\begin{equation}
 \hat H=\sum_{a,b}H_{ab}c_a^\dagger c_b,\qquad H=H^\dagger,
 \label{S:eq:H}
\end{equation}
and linear loss operators
\begin{equation}
 L_\mu=\sum_a\ell_{\mu a}c_a.
 \label{S:eq:jumps}
\end{equation}
The Lindblad equation is
\begin{equation}
 \dot\rho=-i[\hat H,\rho]+\sum_\mu
 \left(L_\mu\rho L_\mu^\dagger-\frac{1}{2}\{L_\mu^\dagger L_\mu,\rho\}\right).
 \label{S:eq:Lindblad}
\end{equation}
Its no-jump Hamiltonian is
\begin{equation}
 \hat K=\hat H-\frac{i}{2}\sum_\mu L_\mu^\dagger L_\mu.
 \label{S:eq:Khat}
\end{equation}
In the one-particle sector,
\begin{equation}
 K=H-\frac{i}{2}M,\qquad
 M_{ab}=\sum_\mu\ell_{\mu a}^*\ell_{\mu b}\succeq0.
 \label{S:eq:passive}
\end{equation}

Let $\Kbar$ be a prescribed complex matrix and consider the family
\begin{equation}
 K_\Gamma=\Kbar-\frac{i\Gamma}{2}\one,
 \qquad \Gamma\in\mathbb R.
 \label{S:eq:family}
\end{equation}
Here $\one$ is the identity on the indicated space: $\one_d$ is the $d\times d$ identity, and $\one_{\rm enc}$ is the identity restricted to the encoded sector. The scalar $\Gamma$ is a signed uniform imaginary translation. If the added uniform loss is required to be nonnegative, its minimum rate is $\max\{0,\Gamma_{\rm CP}\}$. We retain the signed threshold when comparing the numerical and spectral abscissae.
Define the operator imaginary part, using conjugate transpose rather than entrywise complex conjugation,
\begin{equation}
 A\equiv\Iop(\Kbar)=\frac{\Kbar-\Kbar^\dagger}{2i}=A^\dagger.
 \label{S:eq:A}
\end{equation}
Writing $H_0=(\Kbar+\Kbar^\dagger)/2$ gives $\Kbar=H_0+iA$, with both $H_0$ and $A$ Hermitian. Hence $A$ can have complex off-diagonal elements even when $\Kbar$ is real and nonsymmetric. The physical shifted matrix has
\begin{equation}
 \Iop(K_\Gamma)=A-\frac{\Gamma}{2}\one=-\frac12M_\Gamma.
 \label{S:eq:physicalimag}
\end{equation}
Thus $A$ always denotes the operator imaginary part of the centered target, not of the shifted passive generator. The Hermitian and dissipative parts of $K_\Gamma$ are uniquely fixed:
\begin{align}
 H_\Gamma&=\frac{K_\Gamma+K_\Gamma^\dagger}{2}
 =\frac{\Kbar+\Kbar^\dagger}{2},\nonumber\\
 M_\Gamma&=i(K_\Gamma-K_\Gamma^\dagger)
 =\Gamma\one-2A.
 \label{S:eq:HM}
\end{align}

\paragraph{Criterion S1 (minimum uniform scalar shift).}
A passive realization of $K_\Gamma$ exists if and only if
\begin{equation}
 \Gamma\geq\Gamma_{\rm CP}
 =2\lambda_{\max}(A).
 \label{S:eq:GCP}
\end{equation}
General sink embeddings and the numerical-abscissa criterion for passive no-jump realizations were established in Refs.~\cite{Hubisz2021,McDonald2022,Okuma2022}. The proof below fixes the conventions used for reservoir factorization.

\paragraph{Proof.}
Equation~\eqref{S:eq:passive} requires and, by choosing a factorization $M_\Gamma=B^\dagger B$, is implied by $M_\Gamma\succeq0$. From Eq.~\eqref{S:eq:HM}, this is equivalent to
$\Gamma\geq2\lambda_{\max}(A)$. For any such $\Gamma$, the rows of $B$ define valid Lindblad operators and $H_\Gamma$ gives the coherent Hamiltonian. \hfill$\square$

The criterion holds without diagonalizability. A fixed target $K_\Gamma$ uniquely determines the loss matrix, while its factorization into jump operators can be nonunique.

\subsection{Numerical range and spectral threshold}

The numerical range of $\Kbar$ is
\begin{equation}
 W(\Kbar)=\{\langle v|\Kbar|v\rangle:\langle v|v\rangle=1\}.
 \label{S:eq:W}
\end{equation}
Its upper imaginary support is
\begin{equation}
 \max_{z\in W(\Kbar)}\operatorname{Im}z
 =\max_{\|v\|=1}\langle v|A|v\rangle
 =\lambda_{\max}(A).
 \label{S:eq:Wsupport}
\end{equation}
Thus $\Gamma_{\rm CP}/2$ is the signed extremal translation of the numerical range compatible with passivity. When positive, it is the required downward shift.

If $\bar\varepsilon_\alpha$ are the eigenvalues of $\Kbar$, placing the spectrum in the closed decaying half-plane requires
\begin{equation}
 \Gamma\geq\Gamma_{\rm spec}
 =2\max_\alpha\operatorname{Im}\bar\varepsilon_\alpha.
 \label{S:eq:Gspec}
\end{equation}
Every eigenvalue belongs to the numerical range, hence
\begin{equation}
 \Gamma_{\rm spec}\leq\Gamma_{\rm CP}.
 \label{S:eq:inequality}
\end{equation}
Equality holds for normal matrices, because their numerical range is the convex hull of their spectrum. Equality may also occur for some nonnormal matrices, so the excess
\begin{equation}
 \chi_{\rm NH}=\Gamma_{\rm CP}-\Gamma_{\rm spec}
 \label{S:eq:chi}
\end{equation}
is a numerical-abscissa excess associated with nonnormality. It does not characterize normality, since it can also vanish for nonnormal matrices.

For the unnormalized no-jump state $|\psi(t)\rangle=e^{-iK_\Gamma t}|v\rangle$,
\begin{equation}
 \left.\frac{d}{dt}\langle\psi(t)|\psi(t)\rangle\right|_{t=0}
 =2\langle v|A|v\rangle-\Gamma.
 \label{S:eq:normderivative}
\end{equation}
Maximization over normalized $v$ therefore gives
\begin{equation}
 \max_{\|v\|=1}\left.\frac{d}{dt}
 \|e^{-iK_\Gamma t}v\|^2\right|_{t=0}
 =\Gamma_{\rm CP}-\Gamma.
 \label{S:eq:maxnormderivative}
\end{equation}
At $\Gamma=\Gamma_{\rm spec}$ the right-hand side is $\chi_{\rm NH}$. Equality in the spectral condition alone need not imply bounded dynamics when a boundary eigenvalue is defective. Asymptotic stability follows for $\Gamma>\Gamma_{\rm spec}$.

\subsection{Bloch families}

For a translation-invariant lattice with Bloch matrix $\Kbar(k)$, a single uniform shift must satisfy the condition at every momentum:
\begin{align}
 \Gamma_{\rm CP}&=2\max_k
 \lambda_{\max}[\Iop(\Kbar(k))],\nonumber\\
 \Gamma_{\rm spec}&=2\max_{k,\alpha}\operatorname{Im}\bar\varepsilon_\alpha(k).
 \label{S:eq:Blochbounds}
\end{align}
The momentum maximizing the numerical range need not coincide with one maximizing the spectral imaginary part.

\section{A sharp bound on the loss reduction from reservoir support}
\label{S:sec:support}
\label{S:sec:sharp_support}

The comparison-matrix criterion and the support-reducing dephasing maps are
established results of multilevel-coherence and factor-width
theory~\cite{Ringbauer2018,Boman2005,JohnstonCoherence2022}.
In particular, Appendix A.2, Eqs.~(15)-(16), of
Ref.~\cite{Ringbauer2018} gives an explicit map from a $q$-dimensional positive
matrix to sums of positive matrices supported on pairs. We use this
construction to derive a sharp bound on the loss required for passive
reservoir synthesis.

\subsection{Statement and two complementary proofs}

Fix a physical scalar-mode basis with $n\geq q\geq2$ and a prescribed
Hermitian $F$ with $F_{aa}=0$. Let $\Gamma_q$ denote the infimum of the
uniform shifts for which $M_\Gamma=\Gamma\one_n+F=B^\dagger B$ and every
row of $B$ has at most $q$ nonzero entries. Arbitrary complex row amplitudes,
any finite number of rows, and onsite losses are allowed. The pair-supported
comparison must allow every nonzero pair of $F$. There is no independent
constraint on channel count or total coupling power. Then
\begin{equation}
 \boxed{\frac{\Gamma_2}{q-1}\leq\Gamma_q\leq\Gamma_2.}
 \label{S:eq:sharp_support_bound}
\end{equation}
For $F\neq0$ both thresholds are positive. Thus the fractional reduction is
at most $1-1/(q-1)$, and in particular
\begin{equation}
 \boxed{1-\frac{\Gamma_3}{\Gamma_2}\leq\frac12.}
 \label{S:eq:sharp_half_bound}
\end{equation}
For $F=0$, all thresholds vanish and no ratio is defined.

\emph{Constructive proof.} Write $M=\sum_\mu b_\mu^\dagger b_\mu$, where
the row $b_\mu$ has support $I_\mu$ of size $r_\mu\leq q$. Replace this row
by its restriction $b_{\mu,ab}$ to every pair $\{a,b\}\subset I_\mu$, and
add onsite rows with amplitude $\sqrt{q-r_\mu}\,b_{\mu a}$ at every
$a\in I_\mu$. Every off-diagonal entry of the original outer product
appears once, and its diagonal appears $r_\mu-1+q-r_\mu=q-1$ times. Therefore
the new pair-supported factor has Gram matrix
\begin{equation}
 M+(q-2)\operatorname{diag}M=(q-1)\Gamma\one_n+F.
 \label{S:eq:pair_conversion}
\end{equation}
This also covers $r_\mu=1$, for which only the onsite terms remain.
It is the unnormalized $k=2$, $d=q$ dephasing construction of
Ref.~\cite{Ringbauer2018}, applied to each supported row. Taking an
infimizing sequence gives the lower inequality in
Eq.~\eqref{S:eq:sharp_support_bound}. The upper inequality follows from
inclusion of all pair-supported rows among $q$-supported rows. If the pair
conversion temporarily introduces channels on a pair where $F_{ab}=0$,
their summed $2\times2$ block has zero off-diagonal entry and can be replaced
by its onsite diagonal contributions. Thus availability of the nonzero pairs
of $F$ suffices.

\emph{Spectral proof.} Put $W_{aa}=0$ and $W_{ab}=|F_{ab}|$ for $a\neq b$.
The established comparison-matrix criterion gives
$\Gamma_2=\rho(W)$~\cite{Ringbauer2018}. For every real $x\geq0$,
\begin{align}
 x^{T}Wx
 &\leq\sum_\mu\left[
       \left(\sum_a|B_{\mu a}|x_a\right)^2
                   -\sum_a|B_{\mu a}|^2x_a^2\right]\nonumber\\
 &\leq(q-1)\sum_a M_{aa}x_a^2
  =(q-1)\Gamma\|x\|^2.
 \label{S:eq:sharp_perron_proof}
\end{align}
The first inequality is the triangle inequality and the second is
Cauchy-Schwarz on each row's support. Choosing a normalized nonnegative
Perron eigenvector, which exists also for reducible $W$, and then minimizing
over the admissible $\Gamma$ proves the same bound. This argument highlights
why uniform diagonal loss and scalar support, rather than merely spatial
range, are essential to the stated cost ratio.

The finite-dimensional proofs apply directly to any finite periodic chain.
For a translation-invariant finite-range factor, the constructive conversion
can also be performed on every real-space row before Fourier transformation.
It preserves the maximum spatial diameter of that row. Consequently the
same bound applies to the bulk periodic threshold. The counting variable
$q$ refers to physical real-space modes in one row, not to the number of
Bloch orbitals or to the number of periodic channel families.

\subsection{Sharpness, the root chain, and relaxation}

The constant is sharp for every $q$. Let $u\in\mathbb C^q$ have
$|u_a|=1$, and take
\begin{equation}
 F=u u^\dagger-\one_q.
 \label{S:eq:sharp_example}
\end{equation}
Here $\Gamma_q=\Gamma_{\rm CP}=1$, attained by the single row $u^\dagger$,
whereas $W$ is the adjacency matrix of the complete graph and
$\Gamma_2=q-1$. The $q=3$ root chain at
$\phi=\pi/2$, $\gamma=S=t+J$ realizes the sharp factor of two with local
rows, as Eq.~\eqref{S:eq:phase_explicit_rows} shows. Its $50\%$ reduction
therefore saturates Eq.~\eqref{S:eq:sharp_half_bound} over the stated
passive, uniform-loss class. The bound concerns the uniform shift, not
particle survival, reservoir power, or architectures outside this class.

Saturating a static cost inequality does not determine the relaxation gap.
The periodic tuned root model has dark modes. The open-chain construction
at $\gamma=S$ instead has a positive finite-size gap and a reduction slightly
below $50\%$. Retuning the finite-chain coupling to
$\gamma=\|T_L\|_2$ restores equality but also creates a dark sector, as
discussed in \hyperref[S:sec:phase]{the phase-controlled construction below}.
More generally, equality need not force gaplessness: for
Eq.~\eqref{S:eq:sharp_example}, choose $H$ diagonal with distinct real
entries. No eigenvector of $H$ lies in $\ker(u u^\dagger)$, so every
eigenvalue of $H-i u u^\dagger/2$ has strictly negative imaginary part.
A real eigenvalue of a passive matrix would require its eigenvector
to be in the loss kernel and simultaneously an eigenvector of $H$.
The same sharp support cost can therefore coexist with either a dark sector
or strictly decaying finite-dimensional dynamics, depending on the coherent
Hamiltonian.

The scalar ratio need not hold for nonzero diagonal entries of $F$,
additional amplitude or channel-count constraints on the pair realization,
or support defined after a nonlocal basis rotation. In those cases the relevant constrained optimization must be
stated and solved separately. The general matrix conversion
Eq.~\eqref{S:eq:pair_conversion} remains valid, but its diagonal correction
is no longer necessarily a uniform shift of the same target.

\section{Phase-controlled support advantage in the cubic-root chain}
\label{S:sec:phase}

\subsection{Target, phase convention, and exact bulk thresholds}

Fix the physical scalar-mode basis and let $t,J>0$, $\gamma\geq0$,
$S=t+J$, and $h(k)=t+Je^{ik}$. The phase-extended target is
\begin{equation}
 \Kbar_\phi(k)=
 \begin{pmatrix}
  0&0&h^*(k)\\ h(k)&0&0\\0&\gamma e^{i\phi}&0
 \end{pmatrix},\qquad
 \Kbar_\phi^3(k)=\gamma e^{i\phi}|h(k)|^2\one_3.
 \label{S:eq:phase_target}
\end{equation}
The phase modifies the prescribed directed hopping, rather than only its
reservoir factorization. Defining
$D_\phi=\operatorname{diag}(1,e^{-i\phi/3},e^{i\phi/3})$ gives
\begin{equation}
 \Kbar_\phi=e^{i\phi/3}D_\phi\Kbar_0D_\phi^\dagger.
 \label{S:eq:phase_rotation}
\end{equation}
Thus the root multiplets rotate by $\phi/3$, while the diagonal rephasing
preserves all scalar supports. The overall complex factor in
Eq.~\eqref{S:eq:phase_rotation} changes the physical coherent and dissipative
parts. It cannot be removed by a basis rephasing: the product around a directed
three-step cycle retains the phase $\phi$.

With $F_\phi=i(\Kbar_\phi-\Kbar_\phi^\dagger)$ and $r=|h|$, a diagonal unitary
removes the phase of $h$ from its eigenvalue problem. Direct expansion gives
\begin{equation}
 \det(\lambda\one_3-F_\phi)
 =\lambda^3-(2r^2+\gamma^2)\lambda
       -2\gamma r^2\sin\phi.
 \label{S:eq:phase_cubic}
\end{equation}
The minimum eigenvalue of $F_\phi(r)$ is an even concave function of real $r$:
it is a minimum of affine Rayleigh quotients, and changing $r$ to $-r$ is a
diagonal unitary transformation. It is therefore nonincreasing for $r\geq0$.
Since $|h(k)|\leq S$, the bulk optimum is attained at $k=0$:
\begin{align}
 \Gamma_{\rm CP}(\phi,\gamma)&=-\lambda_{\min}[F_\phi(S)],
 \label{S:eq:phase_CP}\\
 \Gamma_2(\gamma)&=\frac{\gamma+\sqrt{\gamma^2+8S^2}}{2}.
 \label{S:eq:phase_pair}
\end{align}
The second expression follows from the phase-independent comparison matrix.
The spectral threshold, which is distinct from both support thresholds, is
\begin{equation}
 \Gamma_{\rm spec}
 =2(\gamma S^2)^{1/3}
 \max_{m=0,1,2}\sin\!\left(\frac{\phi+2\pi m}{3}\right).
 \label{S:eq:phase_spec}
\end{equation}

\subsection{Three-mode attainability throughout the phase diagram}

For any $\Gamma\geq\Gamma_{\rm CP}$, introduce the positive matrix
\begin{equation}
 C_\Gamma=\Gamma\one_3+F_\phi(S),\qquad
 C_\Gamma=R^\dagger R.
 \label{S:eq:triangle_C}
\end{equation}
Each row $R_{\nu a}$ defines two local jump families
\begin{align}
 L^{(t)}_{\nu,j}
 &=\sqrt{\frac{t}{S}}
  (R_{\nu A}c_{A,j}+R_{\nu 1}c_{B_1,j}+R_{\nu 2}c_{B_2,j}),\nonumber\\
 L^{(J)}_{\nu,j}
 &=\sqrt{\frac{J}{S}}
  (R_{\nu A}c_{A,j+1}+R_{\nu 1}c_{B_1,j}+R_{\nu 2}c_{B_2,j}).
 \label{S:eq:triangle_jumps}
\end{align}
Each row has support at most three. The intracell and intercell triangles
contribute $t/S$ and $J/S$ of every diagonal and $B_1$-$B_2$ entry of
$C_\Gamma$, while their $A$-$B$ entries give $t+Je^{ik}$. Their summed Gram
matrix is therefore exactly $\Gamma\one_3+F_\phi(k)$. Consequently,
\begin{equation}
 \boxed{\Gamma_3(\phi,\gamma)=\Gamma_{\rm CP}(\phi,\gamma)}
 \label{S:eq:phase_G3}
\end{equation}
throughout the bulk phase diagram. The construction bounds the number of
physical modes in each row explicitly. A bound on spatial range alone would
not suffice. At a generic threshold $C_\Gamma$ has rank two, so four local
channel families suffice. This count is not generally minimal: the
real-hopping factor derived below uses three.

At $\phi=\pi/2$, Eq.~\eqref{S:eq:phase_cubic} factorizes as
\begin{equation}
 (\lambda+\gamma)(\lambda^2-\gamma\lambda-2S^2),
\end{equation}
and hence
\begin{equation}
 \Gamma_3=\max\!\left\{\gamma,
                  \frac{\sqrt{\gamma^2+8S^2}-\gamma}{2}\right\}.
 \label{S:eq:phase_quadrature}
\end{equation}
Writing $a=(\sqrt{\gamma^2+8S^2}-\gamma)/2$ gives
$\Gamma_2=\gamma+a$ and $\Gamma_3=\max(\gamma,a)$. Thus
$\Gamma_3/\Gamma_2\geq1/2$, with equality precisely at $\gamma=a=S$.
For fixed $\gamma$, the smallest root of Eq.~\eqref{S:eq:phase_cubic} increases
with $\sin\phi$. Away from a repeated root its derivative is
$2\gamma S^2/[3\lambda_{\min}^2-(2S^2+\gamma^2)]\geq0$, and continuity covers
the repeated-root points. Therefore $\phi=\pi/2$ minimizes $\Gamma_3$ and
\begin{equation}
 \boxed{\Gamma_3=S,\quad\Gamma_2=2S,\quad
  1-\Gamma_3/\Gamma_2=\tfrac12}
 \quad(\gamma=S,\ \phi=\pi/2).
 \label{S:eq:phase_half}
\end{equation}
The $50\%$ reduction saturates the sharp bound in
Eq.~\eqref{S:eq:sharp_support_bound} for zero-diagonal $F$, a uniform shift,
and the stated passive linear-loss support classes.

At Eq.~\eqref{S:eq:phase_half}, $C_S$ has rank one and the construction reduces
to two families,
\begin{align}
 L_{t,j}&=\sqrt t\,(c_{A,j}-ic_{B_1,j}+ic_{B_2,j}),\nonumber\\
 L_{J,j}&=\sqrt J\,(c_{A,j+1}-ic_{B_1,j}+ic_{B_2,j}).
 \label{S:eq:phase_explicit_rows}
\end{align}
For $t,J>0$, their Bloch Gram matrix has rank two at generic $k$, proving that
two translation-invariant channel families are necessary as well as sufficient
at this point. Its rank drops to one at $k=0$.

\subsection{Exact open-chain thresholds and boundary reservoirs}

For an open chain of $L$ cells define
$T_L=t\one_L+J U_L$, where $(U_L)_{j,j+1}=1$, and
$\sigma_L=\|T_L\|_2$. In the sublattice-grouped basis,
\begin{equation}
 \Kbar^{(L)}_\phi=
 \begin{pmatrix}
 0&0&T_L^\dagger\\ T_L&0&0\\0&\gamma e^{i\phi}\one_L&0
 \end{pmatrix}.
 \label{S:eq:phase_open_target}
\end{equation}
If $T_L=U\Sigma V^\dagger$, the unitary transformation
$\operatorname{diag}(V,U,U)$ decomposes both the target and $F$ into three-mode
blocks with $r$ equal to the singular values of $T_L$. The comparison matrix
has the same decomposition with all nonzero weights positive. Thus
\begin{align}
 \Gamma_{\rm CP}^{(L)}&=-\lambda_{\min}[F_\phi(\sigma_L)],\nonumber\\
 \Gamma_2^{(L)}&=\frac{\gamma+\sqrt{\gamma^2+8\sigma_L^2}}2.
 \label{S:eq:phase_open_thresholds}
\end{align}

The finite open-chain loss graph admits a perfect elimination order
$A_1,B_{1,1},B_{2,1},A_2,B_{1,2},B_{2,2},\ldots$. At each elimination the
remaining neighbors form a clique of size at most two: for $A_j$ they are
$B_{1,j},B_{2,j}$, and for $B_{1,j}$ they are $B_{2,j},A_{j+1}$.
Eliminating a positive diagonal pivot of a positive-semidefinite matrix
therefore extracts a rank-one Gram row on at most three physical modes, and
the Schur complement remains positive semidefinite on the same remaining
graph. A zero pivot has a zero row by positivity and can be skipped. This
application of established sparse factorization~\cite{Boman2005,Johnston2025}
establishes
\begin{equation}
 \Gamma_3^{(L)}=\Gamma_{\rm CP}^{(L)}.
 \label{S:eq:phase_finite_three}
\end{equation}
The argument also applies to arbitrary complex weights and onsite diagonal
terms on this same open graph. Support within two adjacent cells alone would allow six scalar modes and
would not establish this result.

At $\gamma=S$, $\phi=\pi/2$, one has $\sigma_L<S$ for finite $L$ and $t,J>0$.
Equation~\eqref{S:eq:phase_quadrature} with $S$ replaced only in the $r$ argument
by $\sigma_L$ gives $\Gamma_3^{(L)}=S$. An explicit boundary
construction retains $L_{t,j}$ for $j=1,\ldots,L$, retains $L_{J,j}$ for
$j=1,\ldots,L-1$, and adds
\begin{equation}
 L_{\rm left}=\sqrt J\,c_{A,1},\qquad
 L_{\rm right}=\sqrt J\,(-ic_{B_1,L}+ic_{B_2,L}).
 \label{S:eq:phase_boundaries}
\end{equation}
These $2L+1$ rows satisfy
$B^\dagger B=S\one_{3L}+F^{(L)}_{\pi/2}$ exactly. They give a sufficient,
not necessarily minimal, number of finite-chain rows.

For $t/J=0.5$ and $\gamma/J=1.5$, direct numerical evaluation gives
\begin{center}
\begin{tabular}{rrrrr}
\toprule
$L$ & $\Gamma_3^{(L)}/J$ & $\Gamma_2^{(L)}/J$
 & reduction (\%) & $\Delta_{\mathcal L}^{(L)}/J$\\
\midrule
3 &1.500000&2.807325&46.5684&0.0492725\\
6 &1.500000&2.945727&49.0788&0.0136517\\
24&1.500000&2.996298&49.9382&0.0009259\\
\bottomrule
\end{tabular}
\end{center}
Here $\Gamma_{\rm spec}^{(L)}=(S\sigma_L^2)^{1/3}$ and the gap of the full
loss-only fermionic Liouvillian is
$\Delta_{\mathcal L}^{(L)}=[S-(S\sigma_L^2)^{1/3}]/2$. Rates in a
parity-restricted observable sector must be distinguished from this full
operator-space gap.
If $\gamma$ is retuned from the bulk value $S$ to $\sigma_L$, a finite chain
can also attain exactly $50\%$: then $\Gamma_3^{(L)}=\sigma_L$ and
$\Gamma_2^{(L)}=2\sigma_L$. Its maximum-singular-value sector, however, has
dark eigenmodes and the full gap vanishes. The finite-size table and the
connected benchmark retain $\gamma=S$ and the positive finite-size gap.

\subsection{Nearby targets, dark modes, and the role of nonnormality}

Let $\gamma=S+\delta\gamma$ and $\phi=\pi/2+\delta\phi$. The operator norm of
the perturbation of the $B_1$-$B_2$ block gives a simple nonasymptotic bound,
\begin{equation}
 \Gamma_3(\phi,\gamma)
 \leq S+|\gamma e^{i\phi}-iS|.
 \label{S:eq:phase_tolerance_bound}
\end{equation}
Any parameter pair for which the right-hand side is below
$\Gamma_2(\gamma)$ therefore has a certified strict support advantage. At
$\gamma=S$,
\begin{equation}
 1-\frac{\Gamma_3}{\Gamma_2}
 \geq\frac12-\left|\sin\frac{\delta\phi}{2}\right|,
 \label{S:eq:phase_tolerance_simple}
\end{equation}
which certifies a positive reduction for $|\delta\phi|<\pi/3$.
The exact local expansions are sharper:
\begin{align}
 \frac{\Gamma_3}{S}
 &=1+\frac{|\delta\phi|}{\sqrt3}+O(\delta\phi^2)
 &&(\delta\gamma=0),\nonumber\\
 1-\frac{\Gamma_3}{\Gamma_2}
 &=\frac12-\frac{|\delta\gamma|}{3S}
       +O(\delta\gamma^2/S^2)
 &&(\delta\phi=0).
 \label{S:eq:phase_tolerance_expansion}
\end{align}
The cusp reflects the double minimum eigenvalue of the optimum loss block.
These bounds compare nearby targets, each with its optimal factorization.
Uncalibrated hardware errors at a fixed target are treated in the
common-resource comparison below.

The minimum bulk loss is attained at a boundary of complete positivity. At
$k=0$ the optimum $\Kbar$ is normal, its root eigenvalues are
$S\{e^{i\pi/6},e^{5i\pi/6},e^{3i\pi/2}\}$, and two become real after the
shift $-iS/2$. These are dark modes of the passive Bloch generator, so the
bulk gap closes. Finite open chains have $\sigma_L<S$, a positive gap, and
approach this dark limit as $L$ increases. They do not supply a system-size
independent relaxation gap at optimal loss. Adding a specified positive
uniform margin restores a gap but changes the quoted optimum.

Finally, $\Kbar_\phi$ is normal when the three directed weights have equal
magnitude. Its $k=0$, $\gamma=S$ block therefore shows a strict
support penalty without a numerical-abscissa excess. Away from that block the
connected chain is generally nonnormal. Reservoir support is an independent
resource constraint, and its additional cost must not be attributed solely
to nonnormality.

\subsection{Reproducible verification}

We check 80 seeded random instances
of Eqs.~\eqref{S:eq:phase_rotation}-\eqref{S:eq:phase_open_thresholds}, the
bulk triangle factors, and finite scalar-support-three elimination. The maximum
absolute residual is $9.5\times10^{-12}$ for singular finite factors. The
bulk triangle Gram residual is below $1.1\times10^{-14}$. The special-point
open rows have residual below $5.2\times10^{-15}$. The phase map and tolerance
tables evaluate the analytic thresholds numerically. These checks evaluate the stated mathematical model and do not use experimental measurements.

\section{Certifying collective loss from single-particle measurements}
\label{S:sec:witness}
\label{S:sec:support-witness}

\subsection{A witness for the two-mode support cone}

The dual of the established factor-width-two
cone~\cite{Ringbauer2018,JohnstonCoherence2022} defines a witness accessible
through single-particle loss measurements. A negative witness excludes
every two-mode factorization of the effective loss matrix without
identifying a particular reservoir decomposition.
Fix the physical orbital basis and let $F=F^\dagger$ have zero diagonal.
Write $W_{ab}=|F_{ab}|$ for $a\ne b$, $W_{aa}=0$, and choose a nonnegative
Perron vector $v$ with $Wv=\Gamma_2v$ and $\sum_av_a^2=1$.
Define
\begin{equation}
 Q_{aa}=v_a^2,\qquad
 Q_{ab}=-\frac{F_{ab}}{|F_{ab}|}v_av_b\quad(F_{ab}\ne0),
 \qquad Q_{ab}=0\quad(F_{ab}=0).
 \label{S:eq:witness-Q}
\end{equation}
Every one- or two-dimensional principal submatrix of $Q$ is positive
semidefinite: an edge block has determinant zero, and a nonedge block is
diagonal.  Consequently, if $M=\sum_\mu\ell_\mu^\dagger\ell_\mu$ with each
row $\ell_\mu$ supported on at most two orbitals, then
\begin{equation}
 w(M)\equiv\Tr(QM)=\sum_\mu\ell_\mu Q\ell_\mu^\dagger\ge0.
 \label{S:eq:witness-cone}
\end{equation}
No assumption about translation invariance, identical rates, number of rows,
or a preferred pair decomposition is needed here.  Since $\Tr Q=1$ and
$\Tr(QF)=-v^TWv$, the target family obeys
\begin{equation}
 \boxed{w(M_\Gamma)=\Gamma-\Gamma_2.}
 \label{S:eq:witness-target}
\end{equation}
Thus a negative measured value certifies that the effective Markov loss matrix
requires support greater than two in the specified basis.  It does not assert
uniqueness of the factorization or determine the number of physical reservoirs.

\subsection{A linear number of local preparation settings}

For a normalized one-particle state $|\psi\rangle$, the effective Markov loss
rate is $r(\psi)=\langle\psi|M|\psi\rangle$.  On each edge of $Q$, prepare
\begin{equation}
 |\psi_{ab}\rangle=\frac{|a\rangle+e^{-i\arg Q_{ab}}|b\rangle}{\sqrt2}.
 \label{S:eq:witness-pair-state}
\end{equation}
Then
\begin{align}
 w(M)&=\sum_a\left(Q_{aa}-\sum_{b\ne a}|Q_{ab}|\right)r(a)
       +\sum_{a<b}2|Q_{ab}|r(\psi_{ab}).
 \label{S:eq:witness-readout}
\end{align}
This requires $N+E$ preparations for $N$ orbitals and $E$ nonzero upper-triangle
entries of $Q$, hence a number linear in system size for a bounded-degree
chain.  To check the phase convention, the pair rate contains
$\operatorname{Re}[e^{-i\arg Q_{ab}}M_{ab}]$, and its coefficient in
Eq.~\eqref{S:eq:witness-readout} is exactly
$2\operatorname{Re}[Q_{ab}M_{ba}]$.  A global phase on an initial state is
irrelevant.

For a small device one may instead diagonalize $Q=\sum_\nu q_\nu
|u_\nu\rangle\langle u_\nu|$ and use
$w=\sum_\nu q_\nu r(u_\nu)$.  This is not a local preparation protocol for
an arbitrary large chain.  For the three-mode phase-optimal example, however,
the three states lie on one trimer and
\begin{equation}
 Q=\frac{\one-F/S}{3},\qquad
 \operatorname{spec}Q=\{-1/3,2/3,2/3\},\qquad
 w(M_{\Gamma_3})=-S.
 \label{S:eq:witness-trimer}
\end{equation}
The sum of absolute readout coefficients is $C=5/3$, compared with $C=3$
for the six site/pair settings of Eq.~\eqref{S:eq:witness-readout}.

\subsection{A rigorous correction for an initially empty auxiliary reservoir}

The instantaneous microscopic loss rate at time zero vanishes when all
auxiliaries start empty.  The effective Markov identity
$-\dot P(0)=r(\psi)$ must therefore not be applied to such microscopic data.
We instead count particles emitted by a finite time and use two times
$0<a<b$, with $\Delta=b-a$.  Here $P_{\rm tot}(t)$ is the probability that
the particle remains anywhere in the system plus the auxiliaries, and
$q(t)=1-P_{\rm tot}(t)$.  System-only occupation is a different observable:
its decrease includes coherent transfer into auxiliaries and cannot be used
in the bound below.

Consider the calibrated single-particle model
\begin{equation}
 \dot x=-ihx-i\frac{\sqrt\kappa}{2}B^\dagger y,\qquad
 \dot y=-i\frac{\sqrt\kappa}{2}Bx-\frac\kappa2y,
 \qquad x(0)=\psi,\quad y(0)=0,
 \label{S:eq:witness-micro}
\end{equation}
with $h=h^\dagger$, $M=B^\dagger B$, and common auxiliary linewidth
$\kappa$.  Let $m\ge\|M\|$ and $h_*\ge\|h\|$ be independently calibrated
upper bounds.  They must hold for every admissible realization, including the
two-mode null hypothesis, and must not merely be calculated from the desired
three-mode design.  Contractivity of the full no-jump propagator gives
$\|x(t)\|\le1$.  With $z=i\sqrt\kappa\,y$,
\begin{equation}
 z(t)=\frac\kappa2\int_0^t e^{-\kappa(t-s)/2}Bx(s)\,ds,
 \qquad \dot q(t)=\|z(t)\|^2.
 \label{S:eq:witness-volterra}
\end{equation}
Thus $\|z(t)\|\le\sqrt m$ and
$\|\dot x(t)\|\le D\equiv h_*+m/2$.  Comparing the convolution with
$B\psi$ gives
$\|z(t)-B\psi\|\le\sqrt m\,[e^{-\kappa t/2}+Dt]$.
Integration yields the uniform, nonasymptotic bound
\begin{align}
 \left|\frac{q(b)-q(a)}{\Delta}-r(\psi)\right|
 &\le\mathcal E_\kappa(a,b),\nonumber\\
 \mathcal E_\kappa(a,b)
 &=2m\left[
 \frac{2(e^{-\kappa a/2}-e^{-\kappa b/2})}{\kappa\Delta}
 +\frac{D(a+b)}2\right].
 \label{S:eq:witness-finite-window-bound}
\end{align}
The same result holds for mixed initial states by convexity.  If only a lower
calibration bound $\kappa\ge\kappa_-$ is available, use $\kappa_-$ on the
right-hand side: the first term is the average of a decreasing exponential.
The bound includes both reservoir memory and ordinary evolution of the
prepared state during the measurement window.  No fitted subtraction of the
predicted three-mode signal is used.

If a stationary actual preparation $\rho_\nu$ differs from the nominal
$|u_\nu\rangle$ by trace distance at most $\epsilon_{\rm p}$, then
$|\Tr[M(\rho_\nu-|u_\nu\rangle\langle u_\nu|)]|\le m\epsilon_{\rm p}$.
For readout coefficients $c_\nu$, the total deterministic correction is
therefore
\begin{equation}
 B_{\rm sys}=C[\mathcal E_{\kappa_-}(a,b)+m\epsilon_{\rm p}],
 \qquad C=\sum_\nu|c_\nu|.
 \label{S:eq:witness-systematic}
\end{equation}
This assumes that the same actual preparation is used at both times.  If
each time batch can additionally drift by trace distance $\epsilon_{\rm d}$
from that stationary state, add $2C\epsilon_{\rm d}/\Delta$.
An uncorrected negative finite-time witness alone is not a support certificate.

\subsection{Exact binomial coverage and a synthetic demonstration}

For each state and each time, use an independent ensemble of $n$ trials with
one initially prepared particle.  The number of detected emissions is
binomial.  To include detector calibration, take
$q_{\rm det}=d+(\eta-d)q$ with
$\eta\in[\eta_-,\eta_+]$ and $0\le d\le d_+$.
If an exact Clopper-Pearson interval for $q_{\rm det}$ is $[l,u]$, a valid
interval for $q$ is
\begin{equation}
 \left[\max\!\left(0,\frac{l-d_+}{\eta_+-d_+}\right),
       \min\!\left(1,\frac{u}{\eta_-}\right)\right].
 \label{S:eq:witness-detection}
\end{equation}
Form a rate interval from the lower bound at $b$ minus the upper bound at
$a$, and conversely for its upper endpoint.  For a negative coefficient
$c_\nu$, reverse the interval endpoints when forming the witness sum.
With $J$ probability estimates in a prespecified scan, individual interval
error probabilities $\alpha/J$ give simultaneous coverage at least
$1-\alpha$ by the union bound.  If the resulting finite-window upper endpoint
is $w_+$, the certified criterion is
\begin{equation}
 \boxed{w_++B_{\rm sys}<0.}
 \label{S:eq:witness-certified}
\end{equation}
This is a confidence statement conditional on the stated deterministic
calibration envelopes. Uncertain calibration confidence levels must also be
included in the total error budget in an experiment.

The synthetic demonstration uses
the trimer $S=1$, $\phi=\pi/2$, $\gamma=S$, exact propagation of
Eq.~\eqref{S:eq:witness-micro}, and the three spectral preparations.
The prespecified scan has nine nominal linewidths
$\kappa/S=40,80,160,240,320,480,640,960,1280$,
$a=6/\kappa$, $b=12/\kappa$, and $10^5$ trials per state/time setting.
All 54 probability intervals have simultaneous coverage at least $99\%$.
The assumed calibration envelopes are
\begin{equation}
 m/S=3.05,\quad h_*/S=0.90,\quad
 \kappa_- =0.99\kappa,\quad
 \eta_-=0.979,\quad\eta_+=0.981,\quad
 d_+=10^{-5},\quad\epsilon_{\rm p}=5\times10^{-4}.
 \label{S:eq:witness-calibration}
\end{equation}
The data generator itself has $\eta=0.98$, $d=0$, and exact nominal
preparations.  Keeping the full nonzero calibration envelopes in the
analysis, the point $\kappa/S=480$ gives an upper bound
$(w_++B_{\rm sys})/S=-0.2231$, using $6\times10^5$ trials at that linewidth.
The points at or below $\kappa/S=320$ remain uncertified after the
conservative correction, despite their negative uncorrected values.
The preceding expressions specify the conversion from synthetic counts to confidence intervals, including nuisance-parameter allowances and deterministic bounds. The certificate concerns
effective Markov loss support in the fixed mode basis under the calibrated
microscopic model. It is not a device-independent claim about microscopic
reservoir arity.

\section{A single connected device for support certification and target yield}
\label{S:sec:device}
\label{S:integrated}

We apply support certification and target-state projection to the same
connected-chain device at a fixed linewidth. The results are numerical
predictions from direct system-auxiliary propagation and synthetic
binomial counts.

\subsection{Fixed device, calibration envelope, and error realizations}

The open chain has $L=6$, $t/J=1/2$, $\gamma=S=t+J$, and $\phi=\pi/2$.
Its thresholds are $\Gamma_3/S=1$ and
$\Gamma_2/S=1.9638178232$. Both architectures use $\kappa/S=1920$,
the initial system state $|A_2\rangle$ (zero-based cell indexing), and
the final time $S\tau=1$. The nominal coherent matrix is
$H=(\Kbar+\Kbar^\dagger)/2$ and the finite-device generator is
\begin{equation}
 K_{\rm ext}=
 \begin{pmatrix}H_{\rm act}&G^\dagger\\
 G&D_{\rm aux}-i\kappa\one/2\end{pmatrix},
 \qquad G=\frac{\sqrt\kappa}{2}B_{\rm act}.
 \label{S:eq:integrated-ext}
\end{equation}
There are 13 active collective auxiliaries or 28 active pair auxiliaries,
with a common available bank of 28 and a common per-link ceiling
$|G_{\mu a}|/S\le24.100$. Inactive bank modes are padded explicitly.
The median actual total coupling powers $\sum_{\mu a}|G_{\mu a}|^2/S^2$
are $8631.14$ and $16983.96$, respectively. Total coupling power is reported
as a resource rather than constrained to be equal. The linewidth and
per-link ceiling are fixed throughout the comparison.

We generate 64 paired realizations with independent relative amplitude
and phase errors uniform in $[-0.03,0.03]$ on each allowed nonzero entry of
$B$. The upper-triangular nonzero coherent couplings have the same error
distribution, with conjugation imposed on the lower triangle and the same
$H_{\rm act}$ used by both architectures. A second set uses these same
64 error realizations and adds a common phase
$\theta\in[-0.03,0.03]$ to every $B_2$-sublattice column of $G$, and a
common auxiliary detuning
$D_{\rm aux}=d\one$, $d/\kappa\in[-0.005,0.005]$.
The common phase changes the relative phase of coherent and dissipative
couplings: $H_{\rm act}$ is held fixed, so this is not a simultaneous
basis transformation of the full device. The two error sets are paired
and are not 128 independent disorder samples.

The confidence statements below assume the following deterministic
calibration envelopes, valid also for the competing pair-supported class:
\begin{align}
 \|M_0\|/S&\le4.4,& \|H_{\rm act}\|/S&\le1.05,
 \nonumber\\
 \kappa_{\rm true}&\ge0.99\kappa,&
 \|D_{\rm aux}\|&\le0.005\kappa,
 \nonumber\\
 \eta&\in[0.979,0.981],&d_{\rm click}&\le10^{-5},
 \nonumber\\
 \epsilon_{\rm p}&=5\times10^{-4},&
 \epsilon_{\rm ro}&=10^{-3}.
 \label{S:eq:integrated-calibration}
\end{align}
Here $M_0=4G^\dagger G/\kappa_{\rm true}$ is the resonant-reference
loss Gram matrix, $\epsilon_{\rm p}$ bounds preparation trace distance,
and $\epsilon_{\rm ro}$ bounds the target-readout effect in operator norm
after efficiency correction. The actual generator uses the nominal
linewidth, detector efficiency $0.98$, false-click probability $10^{-6}$,
and exact nominal preparations. Nevertheless, the full nonzero calibration
envelopes are retained in the confidence bounds. The largest realized
$\|M_0\|/S$ and $\|H_{\rm act}\|/S$ are $3.98010$ and $0.861664$.
The envelopes in Eq.~\eqref{S:eq:integrated-calibration} are assumed
calibration capabilities. Experimental confidence statements would also
require an error budget for their calibration.

\subsection{Finite-window bound with auxiliary detuning}

In this derivation $\kappa$ denotes the true linewidth and
$B_{\rm act}=2G/\sqrt\kappa$. The numerical envelopes then use
$\kappa_-=0.99\kappa_{\rm nominal}$.
Let $u(t)$ and $v(t)$ be the unnormalized system and auxiliary amplitudes
for an initially empty auxiliary bank, and define $z(t)=i\sqrt\kappa v(t)$.
Writing $m\ge\|M_0\|$, $h_*\ge\|H_{\rm act}\|$, and
$d_*\ge\|D_{\rm aux}\|$, the auxiliary Volterra solution is
\begin{equation}
 z(t)=\frac\kappa2\int_0^t e^{-\kappa r/2}
 e^{-iD_{\rm aux}r}B_{\rm act}u(t-r)\,dr.
 \label{S:eq:integrated-volterra}
\end{equation}
Passivity gives $\|u(t)\|\le1$, $\|z(t)\|\le\sqrt m$, and
$\|\dot u(t)\|\le h_*+m/2=:D_*$. Moreover,
$\|e^{-iD_{\rm aux}r}-\one\|\le d_*r$, so
\begin{equation}
 \|z(t)-B_{\rm act}\psi\|
 \le\sqrt m\left(e^{-\kappa t/2}+D_*t+\frac{2d_*}{\kappa}\right).
 \label{S:eq:integrated-zbound}
\end{equation}
Since the probability emitted from the entire system-auxiliary space
obeys $\dot q=\|z\|^2$, integration yields
\begin{equation}
 \left|\frac{q(b)-q(a)}{b-a}-\langle\psi|M_0|\psi\rangle\right|
 \le\mathcal E_{\kappa}(a,b)+\frac{4md_*}{\kappa},
 \label{S:eq:integrated-detuned-bound}
\end{equation}
where $\mathcal E_\kappa$ is
Eq.~\eqref{S:eq:witness-finite-window-bound}. This uniform
finite-time bound includes reservoir memory without fitting an adiabatic
model. With an uncertain linewidth, its lower calibration bound replaces
$\kappa$ on the right-hand side. The argument holds for arbitrary
Hermitian $D_{\rm aux}$ within the norm envelope, including auxiliary
mixing, and for mixed preparations by convexity.

For nonzero detuning, the zero-frequency eliminated loss matrix is
\begin{equation}
 M_D=\kappa G^\dagger
 [ (\kappa/2)^2\one+D_{\rm aux}^2]^{-1}G,
 \label{S:eq:integrated-MD}
\end{equation}
which need not equal $M_0$. Functional calculus gives
\begin{equation}
 0\preceq M_0-M_D
 \preceq (2d_*/\kappa)^2M_0,
 \qquad
 \|M_D-M_0\|\le m(2d_*/\kappa)^2.
 \label{S:eq:integrated-conversion}
\end{equation}
Thus a certificate of $\Tr(QM_D)<0$ follows by adding the further
allowance $Cm(2d_*/\kappa_-)^2$, where $C=\|Q\|_1=\sum_\nu|c_\nu|$.
Here this allowance is $0.0008580344S$. The result certifies the support
of the associated zero-frequency Markov loss matrix in the fixed system
basis. It does not assert that finite-bandwidth reduced dynamics is an
exact semigroup or infer the arity of microscopic wiring.

\subsection{Prespecified design and simultaneous finite-count bounds}

For this connected chain the Perron witness has 18 nonzero eigenvalues
and $C=1.9112716283$. We prepare its 18 spectral eigenvectors, using
independent batches at two early times. These preparations are global, unlike the site/pair-state protocol in
\hyperref[S:sec:support-witness]{the preceding support-witness section}.
The numerical calculation constructs their full preparation unitaries.
Before generating any noisy devices or synthetic counts, we compare the
candidate windows $\kappa a\in\{6,8,10,12\}$ and
$\kappa b\in\{14,16,18,20,22,24\}$, using only nominal expected
counts and the conservative error envelope. Minimizing the predicted
upper certificate bound selects
\begin{equation}
 a=8/\kappa,\qquad b=20/\kappa,
 \qquad n=10^6\ \text{trials per state/time}.
 \label{S:eq:integrated-window}
\end{equation}
The window is therefore selected from a nominal design calculation before
any noisy observation.
The resulting memory, detuning, preparation, and conversion allowance
is $0.6273688098S$. No realized model bias is subtracted.

We use the exact binomial and detector-inversion intervals of
\hyperref[S:sec:support-witness]{the preceding support-witness section}, with the total error probability $0.01$
divided among all 5120 reported probability groups. These comprise 36
emission settings and four final-time target/system settings for each
of the 128 devices. The union bound provides at least $99\%$
simultaneous coverage conditional on the calibration envelopes.
A single support certificate costs $3.6\times10^7$ particle preparations.
Including the independent final-time batches costs $4.0\times10^7$
trials per paired-device comparison. The complete synthetic data set
therefore represents $5.12\times10^9$ trials. The bandwidth and sample count quantify the resources required by this
conservative protocol.

All 64 devices with independent link errors pass the certificate, with
largest upper endpoint $w_+/S=-0.135791$. All 64 devices with added common phase and
detuning also pass, with largest upper endpoint $-0.150420$.
Certification succeeds for the sampled devices. This result does not
guarantee success for every error within the calibration envelope.

\subsection{An explicit target-state readout on the same device}

The desired normalized state is
$\psi_{\rm tar}=e^{-i\Kbar\tau}|A_2\rangle/
\|e^{-i\Kbar\tau}|A_2\rangle\|$. After the same device evolves to
$S\tau=1$, the target probability and conditional fidelity are
\begin{equation}
 Y=|\langle\psi_{\rm tar}|u(\tau)\rangle|^2,
 \qquad P_{\rm sys}=\|u(\tau)\|^2,
 \qquad F_{\rm cond}=Y/P_{\rm sys}.
 \label{S:eq:integrated-yield}
\end{equation}
For an operational projection, switch off the reservoir couplings and
apply a number-conserving interferometer $U_{\rm ro}$ satisfying
$U_{\rm ro}\psi_{\rm tar}=|A_0\rangle$, followed by occupation readout
of that mode. Summing occupation over the system modes gives
$P_{\rm sys}$. Auxiliary population is excluded from both probabilities.
An explicit construction removes the last nonzero amplitude successively:
for adjacent components $(a,b)$, apply
\begin{equation}
 \frac{1}{\sqrt{|a|^2+|b|^2}}
 \begin{pmatrix}a^*&b^*\\-b&a\end{pmatrix}.
 \label{S:eq:integrated-givens}
\end{equation}
Seventeen two-mode rotations between adjacent entries of the specified
mode ordering implement the 18-mode projection. The constructed matrices
and their order reproduce unitarity and state transfer to machine
precision. Adjacency in this ordering does not establish physical
locality of the circuit. The construction assumes calibrated coherent
mixing and instantaneous rotations. Finite gate durations are not
simulated. Reservoir couplings must be switched off or sufficiently
suppressed during preparation and readout. The assumed effect bound
$\epsilon_{\rm ro}$ must include switching errors and readout-stage loss,
whose microscopic dynamics is not modeled. The synthetic readout includes an additional coherent
rotation of $5\times10^{-4}$ rad and retains the larger error envelope.
The final-time probability intervals include both preparation error and
the target-readout effect error, in addition to binomial and detector
uncertainty.

For independent link errors, the same certified devices give median
$Y_3/Y_2=2.62455$, with empirical 5-95 percentiles
$[2.50078,2.72647]$. The median target probabilities are $0.545860$
and $0.208173$. Adding common phase and detuning gives median ratio
$2.62431$ and percentiles $[2.50071,2.72601]$.
Across both sets, every conservative target-yield ratio lower endpoint
exceeds $2.4084$. The smallest exact simulated conditional fidelity
of either architecture is $0.998512$. This fidelity is computed from the
propagated state. The conservative finite-count intervals do not
themselves certify $F_{\rm cond}\ge0.99$.
Percentiles summarize the 64 paired realizations in each case and are
not statistical confidence intervals for an unspecified disorder law.

The design, state preparation, readout, and interval construction are fixed by the preceding protocol. The comparison uses
nominal threshold-optimal factors. The controller study below separately
tests a restricted family of finite-time compensations.

\subsection{Acquisition time, control workload, and preparation drift}
\label{S:sec:resource-budget}

The number of trials must be combined with preparation and reset times
to assess an implementation. Let $f_{\rm rep}$ denote the effective rate
of completed trials, including preparation, reservoir switching,
detection, reset, and the operating duty cycle. One support certificate
then requires
\begin{equation}
 T_{\rm cert}=\frac{3.6\times10^7}{f_{\rm rep}},
 \label{S:eq:resource-time}
\end{equation}
in addition to calibration and changes of settings. For example,
$f_{\rm rep}=10^2,10^3,10^4,10^5\,\mathrm{s}^{-1}$ corresponds to
$100$, $10$, $1$, and $0.1$ hours, respectively. Including both
architectures' final-time target and system-occupation batches raises
the count to $4.0\times10^7$ trials per paired comparison. These rates
are requirements, not demonstrated device capabilities.

To make the dynamical scales explicit, take the illustrative conversion
$S/(2\pi)=1\,\mathrm{kHz}$. The fixed dimensionless parameters require
$\kappa/(2\pi)=1.920\,\mathrm{MHz}$ and an available per-link coupling
ceiling $G_{\rm ceil}/(2\pi)=24.100\,\mathrm{kHz}$. The final evolution
time is $\tau=159.155\,\mu\mathrm{s}$. The two emission times are
$a=0.6631\,\mu\mathrm{s}$ and $b=1.6579\,\mu\mathrm{s}$.
Thus the emission window is approximately $0.995\,\mu\mathrm{s}$.
The sum of the programmed evolution times alone is
$18\times10^6(a+b)=41.78\,\mathrm{s}$ per certificate, which excludes
the much larger possible preparation, measurement, and reset overhead.
Rescaling $S$ changes these times and all physical coupling and linewidth
requirements together. It does not leave a fixed apparatus unchanged.

The stored spectral-state construction uses 12-17 two-mode rotations
per input and a total of $5.82\times10^8$ rotations over one certificate.
If each implemented rotation takes $t_{\rm g}$, this construction alone
contributes $5.82\times10^8t_{\rm g}$ to the acquisition time. The target
projection requires a further 17 rotations per target-readout trial.
These counts assume the stated mode ordering and do not establish that
its neighboring entries are directly connected in a physical platform.
Finite-duration controls, routing, switching transients, and loss during
these operations must remain within the preparation and readout envelopes
of Eq.~\eqref{S:eq:integrated-calibration}.

The finite-window witness also gives a quantitative preparation-stability
requirement. Suppose that, for each spectral input, the two stationary
time batches differ from a common reference preparation by trace distance
at most $\epsilon_{\rm d}$, beyond the existing reference error
$\epsilon_{\rm p}$. Since an emission probability is the expectation of
an effect $0\preceq E\preceq\one$, each probability changes by at most
$\epsilon_{\rm d}$. Hence the additional witness allowance is
\begin{equation}
 \frac{B_{\rm drift}}{S}
 =\frac{2C\epsilon_{\rm d}}{S(b-a)}
 =611.6069\,\epsilon_{\rm d}.
 \label{S:eq:resource-drift}
\end{equation}
For the least negative reported upper endpoint,
$w_+/S=-0.1357910$, a value
$\epsilon_{\rm d}<2.2202\times10^{-4}$ retains negativity. A value of
$\epsilon_{\rm d}=1.1101\times10^{-4}$ spends one half of that
margin. This statement applies to the reported data and its fixed
calibration envelopes. It does not guarantee success for all devices
within the fabrication-error range. It assumes stationary Bernoulli
trials within each batch and does not cover arbitrary temporal drift of
the Hamiltonian, reservoirs, or detector. The common-phase and detuning
ensembles sample static devices and therefore do not establish temporal
stability. Interleaved acquisition and separate stability measurements
would be needed to justify these assumptions experimentally.

\subsection{What the target-projection intervals certify}
\label{S:sec:fidelity-budget}

The existing simultaneous intervals yield the conservative conditional
fidelity bound $F_{\rm cond}\ge Y_-/P_+$, using the lower target
probability and upper system-survival probability from separate batches.
Across the reported devices, these bounds range from $0.96196$ to
$0.97646$ for pair loss and from $0.98226$ to $0.98938$ for collective
loss. Thus none of these bounds certifies $F_{\rm cond}\ge0.99$,
although the exact propagated states exceed that value.

This limitation is not solely sampling noise. Setting the sampling
uncertainty to zero while retaining the same detector, preparation, and
readout envelopes gives limiting lower endpoints $0.98705$-$0.98873$
for pair loss and $0.99346$-$0.99435$ for collective loss. These are
limits of the present separate-interval construction, not fundamental
bounds on fidelity estimation. Certifying the pair device above $0.99$
requires a smaller systematic allowance or a different measurement
protocol. Joint mode-resolved target-basis readout could estimate target
occupation conditioned on system survival in a single categorical
experiment, but would require a separate error analysis for
mode-dependent detection and false events. That protocol is not part
of the numerical data reported here.

These resource and stability estimates follow directly from the device parameters and synthetic probability records specified above, without generating additional counts.

\section{Matched finite-bandwidth comparison of reservoir architectures}
\label{S:sec:bandwidth}
\label{S:matchedbenchmark}

The support thresholds constrain an effective Markov generator. To test whether
their difference remains useful at finite reservoir bandwidth, we propagate
both reservoir architectures in the same explicit system-auxiliary model.
All results in this section are numerical predictions under the stated
model and error distribution.

\subsection{Common target and hardware capabilities}

We consider the phase-controlled root chain at $\phi=\pi/2$ and
$\gamma=S=t+J$. For the isolated trimer, $S=1$ and
\begin{equation}
 \Kbar=\begin{pmatrix}0&0&1\\1&0&0\\0&i&0\end{pmatrix},
 \qquad B_3=(1,-i,i),\qquad \Gamma_3=1,\quad\Gamma_2=2.
 \label{S:eq:matchedtrimer}
\end{equation}
This trimer is normal and demonstrates the support cost without invoking
nonnormality. We separately simulate connected open chains with $L=3,6$,
$t/J=1/2$, and $\gamma/J=3/2$. Their centered generators are nonnormal.
The collective factor contains the $L$ intracell rows
$\sqrt t\,(1,-i,i)$ on $(A_j,B_{1j},B_{2j})$, the $L-1$ intercell
rows $\sqrt J\,(1,-i,i)$ on $(A_{j+1},B_{1j},B_{2j})$, and the boundary
rows $\sqrt J c_{A_1}$ and
$\sqrt J(-ic_{B_{1L}}+ic_{B_{2L}})$.
Thus it uses $2L+1$ active auxiliaries and support at most three.
The two-mode factor has one row per nonzero edge of $F=i(\Kbar-\Kbar^\dagger)$.
For a positive Perron vector $v$ of $W=|F|$, a row on edge $a<b$ is
\begin{equation}
 (B_2)_{e,a}=\sqrt{W_{ab}v_b/v_a},\qquad
 (B_2)_{e,b}=F_{ab}/(B_2)_{e,a}.
 \label{S:eq:matchedpair}
\end{equation}
It obeys $B_2^\dagger B_2=\rho(W)\one+F$ and uses $5L-2$ active
auxiliaries. These constructions give sufficient channel counts, without establishing
channel-number minimality.

Both configurations share a bank of $q=\max\{\mathrm{rows}(B_2),
\mathrm{rows}(B_3)\}$ auxiliaries. Inactive rows are padded with zeros.
They have identical auxiliary loss rate $\kappa$, identical ideal
$H=(\Kbar+\Kbar^\dagger)/2$, and couplings
\begin{equation}
 G_r=\frac{\sqrt\kappa}{2}B_r,\qquad
 K_{{\rm full},r}=\begin{pmatrix}
 H&G_r^\dagger\\G_r&-i\kappa\one_q/2
 \end{pmatrix},\qquad r=2,3.
 \label{S:eq:matchedfull}
\end{equation}
Each auxiliary has the jump $\sqrt\kappa\,a_\mu$.
Equation~\eqref{S:eq:matchedfull} is propagated directly by a matrix
exponential, including the initial auxiliary buildup and subsequent
backaction. Adiabatic elimination is used only to identify the nominal
target $H-iB_r^\dagger B_r/2$.

The common available per-link coupling ceiling is chosen as $1.1$ times
the larger nominal $\max|G_{r,\mu a}|$. Every realization in the stated
error sweep respects it. The two schemes share capabilities, while their total squared coupling
strengths differ:
\begin{equation}
 \|G_r\|_{\rm F}^2=\frac{\kappa}{4}\Tr M_r
 =\frac{\kappa n\Gamma_r}{4},\qquad n=3L.
 \label{S:eq:matchedpower}
\end{equation}
For $L=3$ at $\kappa/S=80$, the pair and collective values of
$\|G_r\|_{\rm F}^2/S^2$ are $336.8790$ and $180$. For $L=6$ they
are $706.9744$ and $360$. Their nominal maximum link strengths are
$4.47214S$ and $3.65148S$, respectively, below the same ceiling
$4.91935S$. A physical implementation must also provide the required
phase control and reservoir connectivity. No cost for that control
hardware is included in these numbers.

\subsection{Useful yield and a common conditional-fidelity requirement}

The initial excitation occupies the central $A$ orbital,
$A_{1+\lfloor(L-1)/2\rfloor}$, and all auxiliaries start empty. The
common target state is
\begin{equation}
 |\psi_{\rm tar}(\tau)\rangle=
 \frac{e^{-i\Kbar\tau}|\psi_0\rangle}
 {\|e^{-i\Kbar\tau}|\psi_0\rangle\|}.
 \label{S:eq:matchedtarget}
\end{equation}
Let $u_r(\tau)$ be the system component of
$e^{-iK_{{\rm full},r}\tau}(|\psi_0\rangle,0)^T$. The quantities
compared are
\begin{equation}
 P_{{\rm sys},r}=\|u_r\|^2,\qquad
 Y_r=|\langle\psi_{\rm tar}|u_r\rangle|^2,\qquad
 F_{{\rm cond},r}=Y_r/P_{{\rm sys},r}.
 \label{S:eq:matchedmetrics}
\end{equation}
$Y_r$ is the joint probability of retaining the excitation in the system
and passing the target projection. All loss jumps lead to the vacuum, so
these system-sector probabilities are also obtained from the unconditional
one-excitation Lindblad dynamics. No continuous ideal detector is required
to define them. A final target projection is assumed. Its implementation
and readout error are outside this numerical comparison.
The expected number of independent trials for one successful projection is
$1/Y_r$. At a common evolution time and with the same preparation/readout
overhead, the relative trial cost is therefore $Y_3/Y_2$.
In the Markov limit, $F_{{\rm cond},r}=1$ and
$Y_3/Y_2=e^{(\Gamma_2-\Gamma_3)\tau}$. The finite-bandwidth calculation
tests both the yield and the departure from the desired state.

\subsection{Calibration errors and numerical results}

Every nonzero upper-triangular element of $H$ is multiplied by
$(1+a)e^{i\theta}$, followed by Hermitian conjugation for its partner.
Every nonzero entry of $B_r$ receives the same type of perturbation.
The independent errors $a,\theta$ are uniform on $[-\epsilon,\epsilon]$.
Thus $\epsilon$ bounds relative amplitude error and phase error in radians.
The Hamiltonian error realization is shared between architectures. Their
reservoir-link errors are independent draws from the same distribution.
Nominal zeros remain zero, so support is preserved and complete positivity
is exact in every perturbed system-auxiliary model.

The reproducible sweep uses $64$ paired realizations for each
$\epsilon\in\{0,0.01,0.03,0.05,0.10\}$ and
$\kappa/S\in\{10,20,40,80,160\}$, at $S\tau=1$.
Table~\ref{S:tab:matched} gives the representative $\kappa/S=80$,
$\epsilon=0.03$ point. All $64$ draws at that point satisfy the same
requirement $F_{{\rm cond},2},F_{{\rm cond},3}\geq0.99$ for each
listed system. The reported range spans the empirical 5th-95th percentiles of the
paired yield ratios. It is neither a confidence interval for the population
mean nor a guaranteed error bound.

\begin{table}[t]
\caption{Numerical useful-yield comparison at $S\tau=1$,
$\kappa/S=80$, and $\epsilon=0.03$. The trimer is the normal control.
The $L=3,6$ systems are connected nonnormal open chains. Fidelity entries and gain
are medians over $64$ paired draws.}
\label{S:tab:matched}
\begin{ruledtabular}
\begin{tabular}{ccccc}
$L$ & $Y_3/Y_2$ & 5th-95th percentile & $F_{{\rm cond},2}$ & $F_{{\rm cond},3}$\\
1 & 2.6879 & $[2.5749,2.8573]$ & 0.999432 & 0.999655\\
3 & 2.3790 & $[2.2595,2.4693]$ & 0.999572 & 0.999700\\
6 & 2.6007 & $[2.4814,2.7571]$ & 0.999532 & 0.999707
\end{tabular}
\end{ruledtabular}
\end{table}

These examples show a useful-yield advantage under the common fidelity
requirement and sampled error model. They compare the explicit
threshold-optimal factors of Eq.~\eqref{S:eq:matchedpair}, without optimizing
finite-bandwidth compensation, and therefore do not establish an advantage
over all two-mode controllers. It excludes support crosstalk,
auxiliary detuning noise, and preparation/readout errors. The optimal
loss matrix is singular, and its null space need not be invariant under
$H$ in a connected chain. No size-independent relaxation gap is assumed.
The parameter choices, error distributions, observables, and sample counts above specify this comparison. Numerical methods and data availability are described in the \hyperref[S:sec:numerics]{final section}.

\section{Prescribed projective dynamics and coherent endpoint control}
\label{S:sec:projective-dynamics}

The support thresholds concern reproduction of a prescribed linear generator.
High-fidelity preparation of one state at one time is a different task: a
coherent controller can reproduce a single output even when it does not
reproduce the target map. We make this distinction precise and use a
propagator comparison that includes every input direction.

\subsection{Exact trajectory reproduction}

Let $T(\tau)=\exp(-i\Kbar \tau)$ and let $V(\tau)$ be a differentiable, invertible
propagator on the same $N$ physical modes, with $T(0)=V(0)=\one$. Suppose
that, for every nonzero input $|\psi\rangle$ and every time in an interval starting at zero,
the two propagators give the same normalized output density matrix:
\begin{equation}
 \frac{V(\tau)|\psi\rangle\langle\psi|V^\dagger(\tau)}
 {\langle\psi|V^\dagger(\tau)V(\tau)|\psi\rangle}
 =
 \frac{T(\tau)|\psi\rangle\langle\psi|T^\dagger(\tau)}
 {\langle\psi|T^\dagger(\tau)T(\tau)|\psi\rangle}.
 \label{S:eq:projective-exact}
\end{equation}
Then
\begin{equation}
 V(\tau)=a(\tau)T(\tau),\qquad a(\tau)\ne0,\qquad a(0)=1.
 \label{S:eq:projective-scalar}
\end{equation}
Indeed, $T^{-1}V$ maps every vector onto its own span. Its action on a
basis is diagonal, and applying it to a sum of any two basis vectors forces
their diagonal entries to agree. Thus $T^{-1}V=a\one$. The scalar inherits
differentiability from the propagators, for example through
$a=N^{-1}\Tr(T^{-1}V)$.

For a time-local generator $i\dot V=K(\tau)V$, Eq.~\eqref{S:eq:projective-scalar}
implies
\begin{equation}
 K(\tau)=\Kbar+c(\tau)\one,\qquad
 c(\tau)=i\frac{\dot a(\tau)}{a(\tau)}.
 \label{S:eq:trajectory-generator}
\end{equation}
Assume now that this generator has a Markovian realization
$K(\tau)=H(\tau)-iB^\dagger(\tau)B(\tau)/2$, with at most $q$ nonzero entries in each
row of $B(\tau)$ in the fixed physical basis. Writing
$c(\tau)=\xi(\tau)-i\Gamma(\tau)/2$, where $\xi(\tau)$ is real, gives
\begin{equation}
 B^\dagger(\tau)B(\tau)=\Gamma(\tau)\one+F,\qquad
 \Gamma(\tau)\geq\Gamma_q.
 \label{S:eq:trajectory-loss}
\end{equation}
Consequently,
\begin{equation}
 |a(\tau)|^2=\exp\left[-\int_0^\tau\Gamma(s)\,ds\right]
 \leq e^{-\Gamma_q\tau}.
 \label{S:eq:trajectory-success}
\end{equation}
The static support-optimal realization attains this bound. Hence
time-dependent coherent controls cannot reduce the minimum integrated loss
while reproducing the entire prescribed projective trajectory for every
input under these Markovian support assumptions. Agreement at a single
endpoint, or at finitely many sampled times, does not imply
Eq.~\eqref{S:eq:trajectory-generator}. The argument does not constrain arbitrary
finite-bandwidth embeddings or a change of the physical input and output
basis.

\subsection{A comparison of full propagators}

At a fixed time, let $V$ be the nonzero system-to-system block of the
implemented propagator and let $T=\exp(-i\Kbar \tau)$. Define
\begin{align}
 a&=\frac{\Tr(T^\dagger V)}{\|T\|_F^2},\nonumber\\
 \mathcal E^2&=\min_{b\in\mathbb C}
 \frac{\|V-bT\|_F^2}{\|V\|_F^2}
 =1-\frac{|\Tr(T^\dagger V)|^2}
 {\|T\|_F^2\|V\|_F^2}.
 \label{S:eq:map-residual}
\end{align}
The minimizing scalar is $a$. The complementary quantity
$\mathcal F_{\rm map}=1-\mathcal E^2$ is the overlap of the normalized
pure Choi states of the associated single-Kraus completely positive maps. An admissible scalar attenuation makes $T$ trace-nonincreasing without changing this overlap. It is invariant under
a nonzero scalar multiplying either propagator. It is not a worst-input
state fidelity, and the scalar attenuation must be reported separately.
For a maximally entangled input with an $N$-dimensional reference, the
success probability of the $V$ branch is $\|V\|_F^2/N$ and its conditional
Choi state is $|V\rangle\!\rangle/\|V\|_F$. This gives a process-level
interpretation of Eq.~\eqref{S:eq:map-residual}. The numerical values below
are calculated from propagated matrices, rather than inferred from the
target-state counts.

There is an exact benchmark for all coherent controllers. Let
$\sigma_j$ be the singular values of $T$. For any unitary $U$ acting only on the system modes,
\begin{equation}
 \mathcal E(U,T)\geq
 \epsilon_{\rm coh}(T)
 \equiv\left[1-\frac{(\sum_j\sigma_j)^2}
 {N\sum_j\sigma_j^2}\right]^{1/2}.
 \label{S:eq:coherent-map-floor}
\end{equation}
To prove it, write $T=P\Sigma R^\dagger$. Then
$|\Tr(T^\dagger U)|\leq\sum_j\sigma_j$, with equality at the polar unitary
$U=PR^\dagger$. Substitution into Eq.~\eqref{S:eq:map-residual} proves both
the inequality and its attainability. This is a standard polar-decomposition
best-approximation result~\cite{Higham1986}. The floor vanishes precisely
when $T$ is proportional to a unitary. It allows arbitrary coherent
controls and places no restriction on their range, amplitude, or time
dependence. Restricting those controls cannot lower the floor. It does not
apply to a general trace-decreasing implementation using postselection.

\subsection{Bounds valid for every input}

For $a\ne0$, a complementary operator-norm bound is
\begin{equation}
 E=V/a-T,\qquad
 r=\frac{\|E\|_2}{\sigma_{\min}(T)}.
 \label{S:eq:map-relative-radius}
\end{equation}
If $r<1$, every unit vector satisfies
$\|E\psi\|\leq r\|T\psi\|$. The conditional fidelity with the normalized
target output therefore obeys
\begin{equation}
 F_{\rm cond}(\psi)=
 \frac{|\langle T\psi,V\psi\rangle|^2}
 {\|T\psi\|^2\|V\psi\|^2}\geq1-r^2.
 \label{S:eq:all-input-fidelity}
\end{equation}
For completeness, normalize $x=T\psi/\|T\psi\|$ and write
$(T+E)\psi/\|T\psi\|=(1+\alpha)x+z$, with
$\langle x,z\rangle=0$ and $|\alpha|^2+\|z\|^2\leq r^2$. The inequality
\begin{equation}
 r^2|1+\alpha|^2-(1-r^2)\|z\|^2
 \geq|\alpha+r^2|^2\geq0
\end{equation}
gives Eq.~\eqref{S:eq:all-input-fidelity}. The bound is sharp.

For architecture $j$, let $Y_j(\psi)$ denote projection onto the normalized
target output. Triangle inequalities give
\begin{equation}
 |a_j|^2(1-r_j)^2\|T\psi\|^2
 \leq Y_j(\psi)\leq
 |a_j|^2(1+r_j)^2\|T\psi\|^2,
 \label{S:eq:all-input-yield}
\end{equation}
where the lower bound assumes $r_j<1$. In particular,
\begin{equation}
 \frac{Y_3(\psi)}{Y_2(\psi)}\geq
 \left|\frac{a_3}{a_2}\right|^2
 \left(\frac{1-r_3}{1+r_2}\right)^2
 \label{S:eq:all-input-gain}
\end{equation}
for every input whenever the denominator is nonzero. These guarantees
are consequences of the simulated full propagators and the chosen
auxiliary model. They are distinct from statistical certification using
finite experimental counts. Evaluating them on a time grid certifies the
inputs at those sampled times, not the entire continuum between them.

\subsection{An error-tolerant support lower bound}

The static support bound also has an immediate robustness form. Let $Q$
be the pair-support witness with $\Tr Q=1$ and
$\Tr(QF)=-\Gamma_2$. Suppose a pair-supported loss matrix realizes
\begin{equation}
 M'=\Gamma\one+F+E,\qquad E=E^\dagger,\qquad
 \|E\|_2\leq\varepsilon.
\end{equation}
Positivity of the witness on every pair-supported loss matrix and
trace/operator-norm duality imply
\begin{equation}
 0\leq\Tr(QM')=\Gamma-\Gamma_2+\Tr(QE),\qquad
 \Gamma\geq\Gamma_2-\|Q\|_*\varepsilon,
 \label{S:eq:robust-support-bound}
\end{equation}
where $\|Q\|_*$ is the sum of its singular values. Thus an exact
three-mode realization with shift $\Gamma_3$ still requires less uniform
loss than every pair-supported realization within this generator-error
budget if
$\varepsilon<(\Gamma_2-\Gamma_3)/\|Q\|_*$. The bound permits arbitrary
Hermitian $E$. If the perturbed target retains zero diagonal in $F$, one
may additionally impose $E_{aa}=0$. An operator-norm error $\delta$ in the
no-jump generator gives $\varepsilon\leq2\delta$ in its loss matrix.
This is a robust consequence of the existing witness, not an endpoint
control bound.

\subsection{Finite-bandwidth propagator comparison}

The nominal six-cell device uses the factors and empty auxiliaries specified above. For each $\kappa/S=80,320,1920$, the upper-left block of $\exp(-iK_{\rm ext}\tau)$ is computed at $S\tau=0.05,0.10,\ldots,1.50$. Table~\ref{S:tab:map-benchmark} reports the full-propagator residual and input-uniform bounds at $S\tau=1$. The linewidth scan keeps the desired Markov generator fixed by scaling $G=\sqrt\kappa B/2$. The required coupling capability therefore changes between linewidths. At each linewidth both architectures share the same available bank and link ceiling.

\begin{table}[ht]
\caption{Nominal finite-bandwidth dynamics at $S\tau=1$. Subscripts 2 and 3 denote pair and collective reservoirs. Fidelity and yield bounds follow from the full simulated matrices and hold for every one-particle input. They are not confidence bounds from counting data.}
\label{S:tab:map-benchmark}
\begin{ruledtabular}
\begin{tabular}{cccccc}
$\kappa/S$ & $\mathcal E_2$ & $\mathcal E_3$ & $F_{2,\rm lower}$ & $F_{3,\rm lower}$ & $(Y_3/Y_2)_{\rm lower}$\\
80 & $1.291\times10^{-2}$ & $4.174\times10^{-3}$ & 0.996639 & 0.999590 & 2.233510\\
320 & $3.187\times10^{-3}$ & $1.036\times10^{-3}$ & 0.999797 & 0.999974 & 2.519216\\
1920 & $5.293\times10^{-4}$ & $1.723\times10^{-4}$ & 0.999994 & 0.999999 & 2.604331
\end{tabular}
\end{ruledtabular}
\end{table}

At $\kappa/S=1920$, the upper counterpart of Eq.~\eqref{S:eq:all-input-gain} encloses every input's yield ratio in $[2.604331,2.637841]$ at $S\tau=1$, close to the exact Markov value $2.621687$. Across all 30 sampled times, the fidelity lower bounds remain above $0.999963$ and $0.999993$ for pair and collective devices. At $\kappa/S=80$ the pair bound falls to $0.97843$ by $S\tau=1.5$, so the high fidelity at $S\tau=1$ is not extrapolated to the whole interval.

As diagnostics, 18 basis inputs, all 612 equal-weight two-mode superpositions with relative phases $0,\pi/2,\pi,3\pi/2$, 512 seeded Haar inputs, and the 18 witness eigenvectors are propagated. All 1160 inputs obey the matrix bounds. Their sample minima are not used as proofs of worst-input fidelity. Direct integration with DOP853 agrees with matrix exponentiation to $7.1\times10^{-13}$ in amplitude, and the polar-unitary construction saturates Eq.~\eqref{S:eq:coherent-map-floor} to $2.8\times10^{-15}$.

These calculations use nominal calibrated matrices and contain no disorder draws or synthetic detector counts. The noise ensemble and finite-count support certificate are separate tests. A coherent controller with auxiliary modes and postselection would have a nonunitary system block and is outside the system-unitary benchmark. Likewise, the finite-bandwidth tests compare the stated implementations rather than optimizing over all reservoir designs.

This comparison requires only the specified matrices, input states, and propagation times. No controller optimization is involved. Numerical methods and data availability are described in the \hyperref[S:sec:numerics]{final section}.

\section{Endpoint preparation and restricted scalar controls}
\label{S:sec:controls}
\label{S:controllercheck}

\subsection{An endpoint shortcut using coherent links}

The target in the Letter is a prescribed propagator. To illustrate why preparing one endpoint is insufficient to test it, consider the same central-$A$ input, $S\tau=1$, and target projector, but permit independent amplitudes on the existing 28 coherent links. A static Hermitian matrix with the nominal link phases, no diagonal terms or added edges, and $\max|H_{ab}|/S=0.55$ gives
\begin{equation}
 G=0,\qquad P_{\rm sys}=1,\qquad
 Y=F_{\rm cond}=0.9990247336.
\end{equation}
The matrix has $\|H\|_2/S=0.771653$ and is retained with a fixed verifier. Thus a coherent endpoint shortcut is feasible within the stated per-link ceiling when those amplitudes can be adjusted independently. This is not a claim of global optimality or a realization of $\Kbar$. Its full-propagator mismatch is $0.488507$ at $S\tau=1$. Even the minimum fidelity over the 18 basis inputs is only $0.392915$. The all-unitary lower bound in Eq.~\eqref{S:eq:coherent-map-floor} explains the obstruction to reproducing the full map. The scalar-family comparison below is retained as a limited control example, not evidence of a general endpoint-preparation advantage.

\subsection{Specified scalar family}

Exact realization of a prescribed generator and approximate preparation of
one endpoint state are different optimization tasks. A two-mode competitor
can improve the latter by accepting a small change in conditional dynamics.
We quantify this possibility through a constrained finite-time control
search around the nominal factors. This search uses the
same connected $L=6$ chain, central-$A$ input, target projector, and
$S\tau=1$ as \hyperref[S:matchedbenchmark]{the matched benchmark above}. All results in this section assume noiseless calibrated controls.

For each architecture $r=2,3$, let $G_r^0=\sqrt\kappa B_r/2$ and allow
\begin{equation}
 G_r(\alpha)=\alpha G_r^0,\qquad H_r(\beta)=\beta H,
 \label{S:eq:scalarcontrols}
\end{equation}
with $\alpha\geq0$ and $-1.1\leq\beta\leq1.1$.
The common link ceiling is
$g_{\max}=1.1\max_{r,\mu a}|(G_r^0)_{\mu a}|$. Thus
$\alpha\leq g_{\max}/\max|G_r^0|$ for each architecture. Both have the
same available bank of $28$ auxiliaries and the same coherent-link ceiling
$h_{\max}=1.1\max|H|=0.55S$.
The allowed upper limits on $\alpha$ are $1.1$ for the pair factor and
$1.347219$ for the collective factor. Different limits follow from the
same physical ceiling. Actual coupling power is reported as a resource and is not constrained
to be equal between architectures.

We propagate the full system-auxiliary matrix for every candidate and
maximize $Y_r$ subject to $F_{{\rm cond},r}\geq F_*$ separately for each
architecture. A $33\times45$ rectangular grid covers the entire allowed
$(\alpha,\beta)$ box. Up to eight feasible-grid and nominal starting
points seed constrained sequential quadratic programming. The fidelity constraint includes a $10^{-9}$ interior margin. We report
the largest feasible yield found. The search does not certify a global optimum within the continuous box
or over general finite-bandwidth controls.

\begin{table}[t]
\caption{Finite-time scalar-control search on the connected six-cell chain.
The last three columns give the largest feasible yields found for each
architecture and their ratio. Nominal ratios without rescaling are
$2.60918$ at $\kappa/S=80$ and $2.62123$ at $\kappa/S=1920$.
The rows contain noiseless numerical predictions.}
\label{S:tab:controllercheck}
\begin{ruledtabular}
\begin{tabular}{ccccc}
$\kappa/S$ & $F_*$ & $Y_2$ & $Y_3$ & $Y_3/Y_2$\\
80 & 0.990 & 0.334745 & 0.623275 & 1.86194\\
80 & 0.995 & 0.297789 & 0.601533 & 2.02000\\
1920 & 0.990 & 0.328247 & 0.621830 & 1.89440\\
1920 & 0.995 & 0.290474 & 0.599777 & 2.06482
\end{tabular}
\end{ruledtabular}
\end{table}

At $\kappa/S=1920$ and $F_*=0.99$, the selected pair control is
$(\alpha,\beta)=(0.814462,1.014046)$ and the collective control is
$(0.814750,1.015819)$. The pair yield improves by $57.48\%$ over its
nominal value, whereas the collective yield improves by $13.81\%$.
Consequently the ratio decreases from $2.62123$ to $1.89440$. The collective advantage persists within the tested family, with the
reduced ratio quantifying the comparison under the allowed endpoint controls.
The $F_*=0.995$ rows give a second, stricter fidelity requirement. They
are not a certification of robustness to calibration noise.

These controls preserve auxiliary support and passivity. Their Markov
limit is, however,
\begin{equation}
 K_{{\rm eff},r}(\alpha,\beta)
 =\beta H-\frac{i\alpha^2}{2}(\Gamma_r\one+F),
 \label{S:eq:scalarMarkov}
\end{equation}
which generally changes the prescribed centered generator. They therefore
do not lower its exact support threshold. Conversely, the exact threshold
alone cannot establish optimality for this approximate endpoint task.
The search excludes independently adjustable links, time-dependent
controls, support crosstalk, and optimized auxiliary detunings. It assumes noiseless calibration, whereas the integrated certification
and yield calculation samples the specified error ensemble. Thus the
control search does not establish optimality or robustness under those errors.

At each bandwidth point the capability ceilings are shared and fixed
across the two architectures. Across the two bandwidth points,
$g_{\max}$ scales as $\sqrt\kappa$: it is $4.91935S$ at $\kappa/S=80$
and $24.09979S$ at $\kappa/S=1920$. The comparison therefore fixes the capabilities at each bandwidth but
changes the coupling ceiling between bandwidths. The selected controls, fidelities, and coupling resources are obtained from the grid and constrained search specified above.

\section{Origin-centered complex-chiral symmetry is passive only trivially}
\label{S:sec:affine}

Let $p\geq3$, $\omega=e^{2\pi i/p}$, and let $Z$ be invertible. Suppose a passive finite-dimensional matrix $K=H-iM/2$ obeys
\begin{equation}
 ZKZ^{-1}=\omega K.
 \label{S:eq:chiral}
\end{equation}
For any right eigenpair $Kv=\varepsilon v$, Eq.~\eqref{S:eq:chiral} generates the complete orbit $\{\omega^s\varepsilon\}_{s=0}^{p-1}$. Passivity gives $\operatorname{Im}(\omega^s\varepsilon)\leq0$ for every $s$. On the other hand,
\begin{equation}
 \sum_{s=0}^{p-1}\operatorname{Im}(\omega^s\varepsilon)
 =\operatorname{Im}\left(\varepsilon\sum_{s=0}^{p-1}\omega^s\right)=0.
 \label{S:eq:orbitsum}
\end{equation}
Every term must therefore vanish. A nonzero regular $p$-gon with $p\geq3$ cannot lie on the real axis, so $\varepsilon=0$. Thus the spectrum of $K$ contains only zero. Moreover,
\begin{equation}
 0=\operatorname{Im}\Tr K=-\frac12\Tr M.
 \label{S:eq:traceM}
\end{equation}
Since $M\succeq0$, Eq.~\eqref{S:eq:traceM} implies $M=0$. The matrix $K=H$ is then Hermitian with a vanishing spectrum, hence $K=0$. This proves the origin-centered no-go statement.

A uniform shift gives the admissible affine symmetry
\begin{equation}
 Z\left(K_\Gamma+\frac{i\Gamma}{2}\one\right)Z^{-1}
 =\omega\left(K_\Gamma+\frac{i\Gamma}{2}\one\right).
 \label{S:eq:affine}
\end{equation}
The shift does not modify eigenvectors or Jordan block sizes.

\section{Loss-only Liouvillian spectrum and parity-resolved gaps}
\label{S:sec:liouvillian}

Let $\mathcal H_N$ be the $N$-particle sector and define the operator-space blocks
\begin{equation}
 \mathcal B_{N,M}=\operatorname{span}
 \{|u_N\rangle\langle v_M|:|u_N\rangle\in\mathcal H_N,
 |v_M\rangle\in\mathcal H_M\}.
 \label{S:eq:Bnm}
\end{equation}
The no-jump superoperator
\begin{equation}
 \cL_0\rho=-i(\hat K\rho-\rho\hat K^\dagger)
 \label{S:eq:L0}
\end{equation}
preserves each $\mathcal B_{N,M}$. The recycling term
\begin{equation}
 \mathcal J\rho=\sum_\mu L_\mu\rho L_\mu^\dagger
 \label{S:eq:J}
\end{equation}
maps
\begin{equation}
 \mathcal J:\mathcal B_{N,M}\longrightarrow\mathcal B_{N-1,M-1}.
 \label{S:eq:triangular}
\end{equation}
Ordering operator space by decreasing $N+M$ makes $\cL=\cL_0+\mathcal J$ block triangular. Its eigenvalues are therefore those of the diagonal no-jump blocks~\cite{Torres2014}:
\begin{equation}
 \lambda_{\bm n,\bm m}=-i(E_{\bm n}-E_{\bm m}^*).
 \label{S:eq:lambda}
\end{equation}
For a diagonalizable one-particle matrix with eigenvalues
$\varepsilon_\alpha=\bar\varepsilon_\alpha-i\Gamma/2$, ordinary-fermion many-body energies are
\begin{equation}
 E_{\bm n}=\sum_\alpha n_\alpha\varepsilon_\alpha,
 \qquad n_\alpha\in\{0,1\}.
 \label{S:eq:Fock}
\end{equation}
The decay rate associated with one occupied ket or bra mode is
\begin{equation}
 \kappa_\alpha=\frac{\Gamma}{2}-\operatorname{Im}\bar\varepsilon_\alpha\geq0.
 \label{S:eq:kappa}
\end{equation}
If every $\kappa_\alpha>0$, the vacuum is the unique steady state. In the full complex operator space, the smallest nonzero decay rate is carried by $\mathcal B_{1,0}$ or $\mathcal B_{0,1}$:
\begin{equation}
 \Delta_{\cL}^{\rm full}=\kappa_{\min}
 =\frac{\Gamma-\Gamma_{\rm spec}}{2}.
 \label{S:eq:gap}
\end{equation}
These vacuum-one-fermion coherences are parity odd. Under fermion-parity superselection, physical density matrices and observables occupy the even operator sector. Its slowest block is $\mathcal B_{1,1}$, which gives
\begin{equation}
 \Delta_{\cL}^{\rm even}=2\kappa_{\min}
 =\Gamma-\Gamma_{\rm spec}.
 \label{S:eq:evengap}
\end{equation}
At the signed threshold $\Gamma=\Gamma_{\rm CP}$, which is a nonnegative added loss for the root model,
\begin{equation}
 \Delta_{\cL}^{\rm full}=\frac{\chi_{\rm NH}}{2},
 \qquad
 \Delta_{\cL}^{\rm even}=\chi_{\rm NH}.
 \label{S:eq:optgap}
\end{equation}
If $\chi_{\rm NH}=0$, a dark one-particle mode exists and the vacuum is not the unique steady state. At an exceptional point the eigenvalue multiset remains valid through the characteristic polynomial. The Jordan structure is treated below. The spectral gaps alone do not bound finite-time relaxation uniformly: nonnormal prefactors and Jordan polynomials can delay observable relaxation~\cite{Lee2023}.

\section{Local saturation in one dimension}
\label{S:sec:local}

\subsection{Matrix spectral factorization}

Let
\begin{equation}
 \Kbar(k)=\sum_{r=-R}^{R}\Kbar_r e^{ikr}
 \label{S:eq:finiteK}
\end{equation}
be a $p\times p$ finite-range Bloch matrix. At the optimal shift define
\begin{equation}
 M(k)=\Gamma_{\rm CP}\one_p
 -2\Iop(\Kbar(k))\succeq0.
 \label{S:eq:Mk}
\end{equation}
This is a Hermitian matrix Laurent polynomial of degree at most $R$.

The matrix Fej\'er-Riesz theorem states that a positive-definite Hermitian matrix Laurent polynomial on the unit circle admits a polynomial spectral factor of the same degree~\cite{Ephremidze2009}. The semidefinite case needed here follows by regularization. For $\eta>0$, factor
\begin{equation}
 M_\eta(k)=M(k)+\eta\one_p
 =B_\eta^\dagger(k)B_\eta(k),\qquad
 B_\eta(k)=\sum_{r=0}^{R}B_{\eta,r}e^{ikr}.
 \label{S:eq:regularized}
\end{equation}
The constant Fourier coefficient obeys
\begin{equation}
 \frac{1}{2\pi}\int_{-\pi}^{\pi}\Tr M_\eta(k)\,dk
 =\sum_{r=0}^{R}\|B_{\eta,r}\|_F^2,
 \label{S:eq:boundedcoeff}
\end{equation}
so the coefficients are uniformly bounded as $\eta\to0^+$. A convergent subsequence gives matrices $B_r$ satisfying
\begin{equation}
 M(k)=B^\dagger(k)B(k),\qquad
 B(k)=\sum_{r=0}^{R}B_r e^{ikr}.
 \label{S:eq:factor}
\end{equation}
Thus semidefinite saturation does not increase the degree.

\subsection{Jump operators and locality}

Using the Fourier convention
\begin{equation}
 c_{j,a}=\frac{1}{\sqrt L}\sum_k e^{ikj}c_{k,a},
 \label{S:eq:Fourier}
\end{equation}
the rows of $B(k)$ correspond to
\begin{equation}
 L_{\mu j}=\sum_{r=0}^{R}\sum_{a=1}^{p}
 (B_r)_{\mu a}c_{j+r,a}.
 \label{S:eq:localjumps}
\end{equation}
Their loss matrix is exactly Eq.~\eqref{S:eq:Mk}. Choosing
\begin{equation}
 H(k)=\frac{\Kbar(k)+\Kbar^\dagger(k)}{2}
 \label{S:eq:coherentH}
\end{equation}
gives
\begin{equation}
 H(k)-\frac{i}{2}B^\dagger(k)B(k)
 =\Kbar(k)-\frac{i\Gamma_{\rm CP}}{2}\one_p.
 \label{S:eq:exactlocal}
\end{equation}
Both coherent and dissipative couplings have range at most $R$.

For an open finite chain, regard Eq.~\eqref{S:eq:localjumps} as an infinite-chain family and project the coefficient row $\ell_{\mu j}$ of every jump whose support intersects the sample onto the retained one-particle sites. With $P$ the one-particle spatial projector, including all truncated boundary copies gives
\begin{equation}
 M_{\rm open}=\sum_{\mu,j}(\ell_{\mu j}P)^\dagger(\ell_{\mu j}P)
 =P M_\infty P,
 \label{S:eq:boundary}
\end{equation}
which is the principal open-boundary loss matrix. Hence the bulk-optimal uniform shift remains an exact, same-range realization at the edges. It is a sufficient bound for each finite sample. Boundary removal may lower that sample's own minimum scalar shift.

If the factor has $q$ rows, then $\operatorname{rank}M(k)\leq q$ for every $k$. Therefore
\begin{equation}
 q\geq\max_k\operatorname{rank}M(k).
 \label{S:eq:rankbound}
\end{equation}
This lower bound is attainable also for symbols that are rank deficient at every momentum. Let $q_*=\max_k\operatorname{rank}M(k)$. The rank equals $q_*$ except at isolated momenta: a nonzero Laurent-polynomial minor has only finitely many zeros on the unit circle. Apply the rank-deficient spectral factorization theorem~\cite{Ephremidze2015} to the positive-semidefinite symbol $M(k)^T$. It provides a $p\times q_*$ polynomial $C(k)$ of degree at most $R$ with $M(k)^T=C(k)C(k)^\dagger$. Transposing gives
\begin{equation}
 M(k)=\overline{C(k)}C(k)^T=B(k)^\dagger B(k),
 \qquad B(k)=C(k)^T\in\mathbb C^{q_*\times p}.
 \label{S:eq:rectangularfactor}
\end{equation}
Hence $q_{\min}=q_*$ with no range overhead. The case $q_*=0$ requires no jumps. The count refers to independent Markovian linear-loss channels per chosen unit cell, without auxiliary memory or time dependence. The polynomial degree is an existence guarantee, not a guarantee of smooth calibration at rank-changing parameter values: minimal rank-deficient factors need not depend continuously on the symbol~\cite{Ephremidze2015}. Locality-preserving factorizations have also been used in the distinct setting of phonon lattice dynamics~\cite{Li2025}.

\subsection{Finite and disordered open chains}

The finite sample need not inherit the bulk threshold. For an arbitrary open, block-banded target $\Kbar_N$ with cell bandwidth $R$, set
\begin{equation}
 \Gamma_N=2\lambda_{\max}(\Iop(\Kbar_N)),\qquad
 M_N=\Gamma_N\one-2\Iop(\Kbar_N)\succeq0.
 \label{S:eq:finiteoptimal}
\end{equation}
There is a factor $M_N=B_N^\dagger B_N$ with exactly $\operatorname{rank}M_N$ nonzero rows, each supported on at most $R+1$ consecutive cells. This is the banded factor-width-rank result of Johnston, Moein, and Plosker~\cite{Johnston2025}, translated into loss-channel language. Here the channel count is the total for the finite sample, rather than a count per unit cell.

A constructive proof uses positive-semidefinite elimination from the leftmost cell. Partition the remaining matrix and diagonalize its first diagonal block on its positive support:
\begin{equation}
 M=\begin{pmatrix}D&C\\ C^\dagger&E\end{pmatrix},\qquad
 D=U_r\Lambda_r U_r^\dagger,\qquad r=\operatorname{rank}D.
 \label{S:eq:finitepartition}
\end{equation}
Positivity implies $\operatorname{ran}C\subseteq\operatorname{ran}D$ and the generalized Schur complement $S=E-C^\dagger D^+C\succeq0$. Define
\begin{equation}
 F=\left[\Lambda_r^{1/2}U_r^\dagger\quad
          \Lambda_r^{-1/2}U_r^\dagger C\right],\qquad
 M=F^\dagger F+\operatorname{diag}(0,S).
 \label{S:eq:finiteelimination}
\end{equation}
The $r$ rows of $F$ reach only the next $R$ cells. The Schur update couples only these same cells to one another, whose pairwise separations are at most $R-1$. It creates no larger cell bandwidth. An invertible block congruence gives $\operatorname{rank}M=r+\operatorname{rank}S$. Iteration recovers the cited simultaneous locality and factor-width-rank statement, including singular diagonal blocks. A zero block has $C=0$ and contributes no row. Its physical consequence is a band-preserving reservoir factor for arbitrary finite-range disorder at the finite sample's own threshold.

\section{Real-phase baseline: optimal cubic-root SSH realization}
\label{S:sec:realphase}

This section fixes $\phi=0$. The phase-controlled generalization and the maximum 50\% reduction are derived in Eqs.~\eqref{S:eq:phase_G3}-\eqref{S:eq:phase_half}. The 13.40\% maximum below applies only to the real-phase baseline.

\subsection{Complete-positivity and spectral bounds}

For $t,J,\gamma>0$, consider
\begin{equation}
 \Kbar_3(k)=
 \begin{pmatrix}
 0&0&h^*(k)\\
 h(k)&0&0\\
 0&\gamma&0
 \end{pmatrix},\qquad h(k)=t+Je^{ik},
 \label{S:eq:p3}
\end{equation}
for which
\begin{equation}
 \det\left[a\one_3-\Iop(\Kbar_3(k))\right]
 =a\left[a^2-\frac{\gamma^2+2|h(k)|^2}{4}\right].
 \label{S:eq:Achar}
\end{equation}
The eigenvalues of $A(k)$ are therefore
\begin{equation}
 0,\qquad \pm\frac12\sqrt{\gamma^2+2|h(k)|^2}.
 \label{S:eq:Aeigs}
\end{equation}
Since $\max_k|h(k)|=S=t+J$ for $t,J\geq0$,
\begin{equation}
 \Gamma_{\rm CP}=\sqrt{2S^2+\gamma^2}.
 \label{S:eq:GCPp3}
\end{equation}
The centered eigenvalues obey
\begin{equation}
 \bar\varepsilon_s(k)=\omega^s
 \left[\gamma|h(k)|^2\right]^{1/3},
 \label{S:eq:rootbands}
\end{equation}
so
\begin{equation}
 \Gamma_{\rm spec}=\sqrt3\left(\gamma S^2\right)^{1/3}.
 \label{S:eq:Gspecp3}
\end{equation}
Writing $x=\gamma/S$, equality of Eqs.~\eqref{S:eq:GCPp3} and \eqref{S:eq:Gspecp3} is equivalent to
\begin{equation}
 x^2+2=3x^{2/3}
 \quad\Longleftrightarrow\quad
 (x^{2/3}-1)^2(x^{2/3}+2)=0.
 \label{S:eq:equality}
\end{equation}
Thus equality occurs only at $\gamma=S$. At the maximizing momentum $k=0$, all three directed cycle weights then have equal magnitude and $\Kbar_3(0)$ is normal.

The bond-by-bond construction assigns one collective reservoir to every directed matrix element and then balances the onsite rates. It requires
\begin{equation}
 \Gamma_{\rm bond}=\max\{2(t+J),t+J+\gamma\}.
 \label{S:eq:Gedge}
\end{equation}
This simple sufficient bound is generally above the optimum.

\subsection{Exact optimum under two-mode jump support}

The number of physical modes addressed by one jump is distinct from both its spatial diameter and the total number of channels. Let
\begin{equation}
 F=i(\Kbar-\Kbar^\dagger)=F^\dagger,
 \qquad M_\Gamma=\Gamma\one+F.
 \label{S:eq:Fdefinition}
\end{equation}
A factor $M_\Gamma=B^\dagger B$ is \emph{pair supported} when every row of $B$ has at most two nonzero entries in the fixed scalar-mode basis. Support-one onsite rows are allowed. This is factor width at most two, equivalently scaled diagonal dominance in the established matrix terminology~\cite{Boman2005,Johnston2025}. After trace normalization, it is also the established coherence-number-two condition~\cite{Ringbauer2018,JohnstonCoherence2022}. The following derivation translates that known comparison-matrix criterion into uniform reservoir loss.

For a general Hermitian $F$, define the real symmetric Metzler comparison matrix
\begin{equation}
 {\cal C}(F)_{aa}=-F_{aa},\qquad
 {\cal C}(F)_{ab}=|F_{ab}|\quad(a\ne b).
 \label{S:eq:comparison}
\end{equation}
The minimum signed uniform translation compatible with pair-supported loss is
\begin{equation}
 \boxed{\Gamma_2^{\rm signed}=\lambda_{\max}[{\cal C}(F)]}.
 \label{S:eq:Gamma2general}
\end{equation}
If only nonnegative uniform loss may be added, the physical minimum is $\max\{0,\Gamma_2^{\rm signed}\}$. When $F_{aa}=0$, ${\cal C}(F)=W\geq0$ and Eq.~\eqref{S:eq:Gamma2general} reduces to
\begin{equation}
 \boxed{\Gamma_2=\rho(W)},\qquad
 W_{ab}=|F_{ab}|\ (a\ne b),\quad W_{aa}=0.
 \label{S:eq:Gamma2rho}
\end{equation}

To prove necessity, collect all rows supported on $\{a,b\}$ and denote their total endpoint weights by $q_{ab}$ and $q_{ba}$. Cauchy-Schwarz gives $|F_{ab}|\leq\sqrt{q_{ab}q_{ba}}$. If $s_a\geq0$ is the onsite contribution, then for every real vector $x\geq0$,
\begin{align}
 \sum_a(\Gamma+F_{aa})x_a^2
 &=\sum_as_ax_a^2+\sum_{a<b}(q_{ab}x_a^2+q_{ba}x_b^2)\nonumber\\
 &\geq2\sum_{a<b}|F_{ab}|x_ax_b.
 \label{S:eq:pairlower}
\end{align}
Thus $\Gamma\|x\|^2\geq x^{\mathsf T}{\cal C}(F)x$. The largest-eigenvalue eigenvector of a symmetric Metzler matrix can be chosen nonnegative, proving the lower bound in Eq.~\eqref{S:eq:Gamma2general}. The argument permits arbitrary phases, repeated pair channels, cancellations, and onsite rows.

For sufficiency, decompose the graph $F_{ab}\ne0$ into connected components $C_\alpha$. Let $\lambda_\alpha$ and $v^{(\alpha)}>0$ be the Perron eigenpair of ${\cal C}(F)_{C_\alpha}$ and set $\Gamma_*=\max_\alpha\lambda_\alpha$. On every nonzero edge $a<b$ of $C_\alpha$, use
\begin{equation}
 b^{(ab)}_a=\sqrt{|F_{ab}|\frac{v_b}{v_a}},\qquad
 b^{(ab)}_b=e^{i\arg F_{ab}}
 \sqrt{|F_{ab}|\frac{v_a}{v_b}}.
 \label{S:eq:pairPerronrow}
\end{equation}
All other entries vanish. Its off-diagonal Gram element is $F_{ab}$ and the edge rows contribute
\begin{equation}
 \sum_{b\ne a}|F_{ab}|\frac{v_b}{v_a}=\lambda_\alpha+F_{aa}
 \label{S:eq:pairPerrondiag}
\end{equation}
at vertex $a$. Adding the onsite row $\sqrt{\Gamma_*-\lambda_\alpha}\,e_a$ on each component gives $B^\dagger B=\Gamma_*\one+F$. Isolated vertices correspond to $\lambda_\alpha=-F_{aa}$. This proves attainment, including disconnected graphs and prescribed onsite terms.

For a periodic cubic-root chain with $N\geq2$ cells, $t,J,\gamma>0$, and $S=t+J$, the comparison graph is connected. Its translation-invariant Perron vector may be chosen as
\begin{equation}
 v_{A_j}=\frac{2S}{\Gamma_2},\qquad
 v_{B_{1j}}=v_{B_{2j}}=1,
 \label{S:eq:p3Perronvector}
\end{equation}
and its three-orbital quotient is
\begin{equation}
 W_{\rm quot}=\begin{pmatrix}0&S&S\\S&0&\gamma\\S&\gamma&0\end{pmatrix}.
 \label{S:eq:p3quotient}
\end{equation}
Consequently,
\begin{equation}
 \boxed{\Gamma_2=\frac{\gamma+\sqrt{\gamma^2+8S^2}}{2}}.
 \label{S:eq:p3Gamma2}
\end{equation}
With $p=\sqrt{\Gamma_2/(2S)}$ and $q=p^{-1}$, one attaining pair-supported factor is
\begin{align}
 \ell_{1j}&=\sqrt t\,(pA_j-iqB_{1j}),&
 \ell_{2j}&=\sqrt J\,(pA_{j+1}-iqB_{1j}),\nonumber\\
 \ell_{3j}&=\sqrt t\,(qB_{2j}-ipA_j),&
 \ell_{4j}&=\sqrt J\,(qB_{2j}-ipA_{j+1}),\nonumber\\
 \ell_{5j}&=\sqrt\gamma\,(B_{1j}-iB_{2j}).
 \label{S:eq:p3fiveoptimal}
\end{align}
The diagonal Gram entries are $2Sp^2=\Gamma_2$ on $A$ and $Sq^2+\gamma=\Gamma_2$ on $B_1,B_2$. There are $5N$ nonzero unordered physical pairs, while one pair-supported row can cover at most one pair. Equation~\eqref{S:eq:p3fiveoptimal} therefore attains both the exact loss threshold and the lower bound of five channel families per cell, without assuming translation symmetry in the competing architecture.

The three-mode factor below attains
$\Gamma_3=\Gamma_{\rm CP}=\sqrt{\gamma^2+2S^2}$. Since $\Gamma_2^2=\gamma\Gamma_2+2S^2$ and $\Gamma_2>\gamma$,
\begin{equation}
 \boxed{\Gamma_3<\Gamma_2\leq\Gamma_{\rm bond}}
 \qquad(S,\gamma>0).
 \label{S:eq:exactresourcehierarchy}
\end{equation}
At $t=0.5$, $J=\gamma=1$, the rates are $\Gamma_3=2.345207879912$, $\Gamma_2=2.679449471770$, and $\Gamma_{\rm bond}=3$. Thus the exact reduction relative to the optimal two-mode class is $12.4742637\%$. Writing $y=\gamma/\Gamma_2$ gives $(\Gamma_3/\Gamma_2)^2=1-y+y^2\geq3/4$, so the maximum reduction is
\begin{equation}
 1-\frac{\sqrt3}{2}=13.3974596\%,
 \label{S:eq:maxsupportsaving}
\end{equation}
attained at $\gamma=S$. At that point $\Gamma_{\rm spec}=\Gamma_3$ but $\Gamma_2=2S$, separating support cost from the numerical-abscissa excess. For an open finite chain the threshold is $\Gamma_{2,L}=\lambda_{\max}[{\cal C}(F_L)]$. The periodic bulk formula does not generally apply.

\subsection{Bond-by-bond reference construction and affine root geometry}

\begin{figure}[t]
 \centering
 \includegraphics[width=\textwidth]{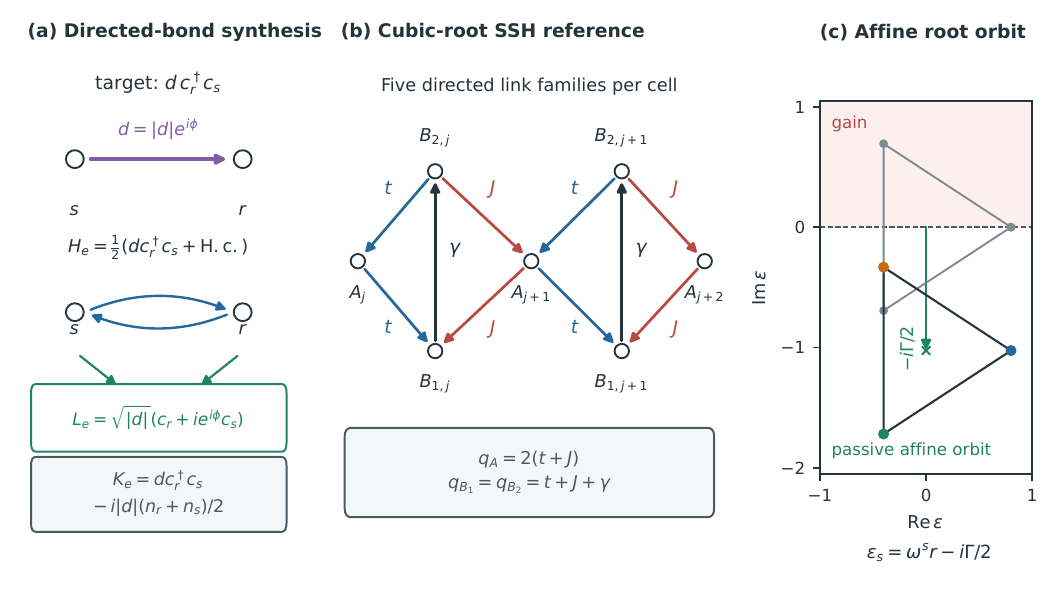}
 \caption{\label{S:fig:bond}
 Reference construction and affine geometry. (a) A reciprocal coherent half-hop together with a phase-biased collective loss synthesizes one directed matrix element of the no-jump Hamiltonian and adds equal endpoint losses. (b) The cubic-root SSH target contains five directed links per cell. Onsite balancing losses are not drawn. The weighted degrees in Eq.~\eqref{S:eq:p3degrees} give the sufficient bond-by-bond shift. (c) An origin-centered $p=3$ root orbit necessarily enters the gain half-plane, whereas a uniform shift $-i\Gamma/2$ produces a passive affine orbit without changing eigenvectors or Jordan structure.}
\end{figure}

Figure~\ref{S:fig:bond} illustrates the elementary reservoir-engineering identity underlying the sufficient bound in Eq.~\eqref{S:eq:Gedge}. Write a target directed matrix element as $d=|d|e^{i\phi}$ and define
\begin{align}
 H_e&=\frac12\left(d\,c_r^\dagger c_s+d^*c_s^\dagger c_r\right),\nonumber\\
 L_e&=\sqrt{|d|}\left(c_r+i e^{i\phi}c_s\right).
 \label{S:eq:directedpair}
\end{align}
A direct expansion gives
\begin{equation}
 H_e-\frac{i}{2}L_e^\dagger L_e
 =d\,c_r^\dagger c_s-\frac{i|d|}{2}(n_r+n_s),
 \label{S:eq:directedidentity}
\end{equation}
so the reverse coherent hop cancels while equal endpoint losses remain. For a finite directed matrix $\Kbar=\sum_{r\neq s}d_{rs}c_r^\dagger c_s$, applying Eq.~\eqref{S:eq:directedidentity} to every ordered edge produces onsite rates
\begin{equation}
 q_a=\sum_{b\neq a}\left(|d_{ab}|+|d_{ba}|\right).
 \label{S:eq:weighteddegree}
\end{equation}
For any $\Gamma\geq\max_a q_a$, the balancing jumps
\begin{equation}
 L_a^{(0)}=\sqrt{\Gamma-q_a}\,c_a
 \label{S:eq:balancingjump}
\end{equation}
complete the exact passive realization $K=\Kbar-i\Gamma\one/2$.

For the cubic-root SSH cycle, the five directed terms in each bulk cell are
\begin{equation}
 tB_{1j}^\dagger A_j+JB_{1j}^\dagger A_{j+1}
 +tA_j^\dagger B_{2j}+JA_{j+1}^\dagger B_{2j}
 +\gamma B_{2j}^\dagger B_{1j}.
 \label{S:eq:fivedirected}
\end{equation}
Their weighted degrees are
\begin{equation}
 q_A=2(t+J),\qquad q_{B_1}=q_{B_2}=t+J+\gamma,
 \label{S:eq:p3degrees}
\end{equation}
which reproduces Eq.~\eqref{S:eq:Gedge}. At the normal point $\gamma=t+J$ the reference construction uses $\Gamma=2(t+J)$. The same uniform shift translates the forbidden origin-centered root polygon into a passive affine orbit. This construction generally requires more loss and more channels than the optimal three-channel factor derived next.

\Needspace{0.30\textheight}
\subsection{Closed-form degree-one factor}

Let
\begin{equation}
 G=\Gamma_{\rm CP}=\sqrt{2S^2+\gamma^2},\qquad
 D=S(t+2J)+\gamma^2.
 \label{S:eq:GD}
\end{equation}
The optimal loss matrix can be written, with $z=e^{ik}$,
\begin{equation}
 M(z)=M_0+M_1z+M_1^\dagger z^{-1},
 \label{S:eq:Mlaurent}
\end{equation}
where
\begin{equation}
 M_0=
 \begin{pmatrix}
 G&-it&it\\ it&G&-i\gamma\\-it&i\gamma&G
 \end{pmatrix},\qquad
 M_1=
 \begin{pmatrix}
 0&0&0\\ iJ&0&0\\-iJ&0&0
 \end{pmatrix}.
 \label{S:eq:M01}
\end{equation}
Define
\begin{align}
 \alpha&=\sqrt{\frac{Gt}{S}},&
 \beta&=\sqrt{\frac{tS}{G}},&
 d&=\sqrt{\frac{D}{G}},\nonumber\\
 e&=\sqrt{\frac{G}{D}}
 \left(\frac{tS}{G}-i\gamma\right),&
 f&=\sqrt{\frac{2JSG}{D}},\nonumber\\
 u&=iJ\sqrt{\frac{G}{D}},&
 v&=\sqrt{\frac{JG}{2SD}}(\gamma-iG).
 \label{S:eq:coefficients}
\end{align}
Then
\begin{equation}
 B_0=
 \begin{pmatrix}
 \alpha&-i\beta&i\beta\\
 0&d&e\\
 0&0&f
 \end{pmatrix},\qquad
 B_1=
 \begin{pmatrix}
 0&0&0\\
 u&0&0\\
 v&0&0
 \end{pmatrix}
 \label{S:eq:B01}
\end{equation}
satisfy the three identities
\begin{equation}
 B_0^\dagger B_0+B_1^\dagger B_1=M_0,
 \qquad B_0^\dagger B_1=M_1,
 \qquad B_1^\dagger B_0=M_1^\dagger.
 \label{S:eq:factorcheck}
\end{equation}
Equations~\eqref{S:eq:Mlaurent}-\eqref{S:eq:factorcheck} prove
$M(k)=B^\dagger(k)B(k)$ with $B(k)=B_0+B_1e^{ik}$.

The corresponding three jumps are
\begin{align}
 L_{1j}&=\alpha A_j-i\beta B_{1j}+i\beta B_{2j},\nonumber\\
 L_{2j}&=uA_{j+1}+dB_{1j}+eB_{2j},\nonumber\\
 L_{3j}&=vA_{j+1}+fB_{2j}.
 \label{S:eq:threejumps}
\end{align}
For $t,J>0$, the loss matrix is singular only at $k=0$ and full rank at generic momenta, so three channels are necessary by Eq.~\eqref{S:eq:rankbound}. At $t=0$, $\alpha=\beta=0$ and the first row vanishes. The symbol has rank two for every momentum and two channels suffice in the exact trimer limit.

For the representative real-phase parameters,
\begin{equation}
 t=0.5,\qquad J=\gamma=1,
 \label{S:eq:params}
\end{equation}
we obtain
\begin{align}
 \Gamma_{\rm spec}&=2.269629\ldots,\nonumber\\
 \Gamma_3=\Gamma_{\rm CP}&=2.345208\ldots,\nonumber\\
 \Gamma_2&=2.679449\ldots,\nonumber\\
 \Gamma_{\rm bond}&=3,
 \label{S:eq:numbers}
\end{align}
and
\begin{equation}
 \kappa_{\min}^{\rm opt}=0.037790\ldots,\qquad
 \Delta_{\cL}^{\rm even}=0.075579\ldots.
 \label{S:eq:gapnumber}
\end{equation}
The three-mode factor lowers the required uniform loss by $12.47\%$ relative to the exact optimum over all two-mode jumps. The larger $21.8\%$ reduction is only relative to the nonoptimal equal-endpoint construction of Eq.~\eqref{S:eq:directedpair}.

\subsection{Finite-size thresholds and conditional yield}

For $L$ open cells, let $T=t\one_L+J S_L$, where $S_L$ has ones on its first superdiagonal. In grouped orbital order,
\begin{equation}
 \Kbar_N=\begin{pmatrix}0&0&T^\dagger\\T&0&0\\0&\gamma\one_L&0\end{pmatrix}.
 \label{S:eq:finitecubic}
\end{equation}
A singular-value decomposition $T=U\Sigma V^\dagger$, with the unitary change of basis $\operatorname{diag}(V,U,U)$, reduces this matrix to cubic blocks with $h=s_j$, the singular values of $T$. Therefore
\begin{equation}
 \Gamma_{\rm CP}^{(L)}=\sqrt{\gamma^2+2\|T\|_2^2},\qquad
 \Gamma_{\rm spec}^{(L)}=\sqrt3\,(\gamma\|T\|_2^2)^{1/3}.
 \label{S:eq:finitethresholds}
\end{equation}
For $t,J>0$, $\|T\|_2<t+J$, so the finite-sample unrestricted threshold is strictly below the bulk value. The largest singular value is simple and $\operatorname{rank}M_N=3L-1$. The scalar perfect-elimination proof leading to Eq.~\eqref{S:eq:phase_finite_three} gives exactly $3L-1$ nonzero rows, each supported on at most three physical modes in two adjacent cells. At $t=0$ and $L\geq2$, the largest singular value $J$ has multiplicity $L-1$, giving $2L+1$ rows. This scalar-support conclusion uses the triangle graph, not merely a two-cell range bound. The finite pair-supported threshold is instead
\begin{equation}
 \Gamma_2^{(L)}=\lambda_{\max}[{\cal C}(F_L)],
 \qquad F_L=i(\Kbar_L-\Kbar_L^\dagger).
 \label{S:eq:finiteGamma2}
\end{equation}

Uniform-shift savings have a direct trajectory interpretation. For an initial state of fixed total particle number $n$, the no-jump generator is
$\hat K_\Gamma=\hat{\Kbar}-i\Gamma\hat N/2$. Since $[\hat{\Kbar},\hat N]=0$,
\begin{equation}
 P_{\rm nj}^{(\Gamma)}(\tau)
 =e^{-n\Gamma\tau}\operatorname{Tr}\left[
 e^{-i\hat{\Kbar}\tau}\rho_n e^{i\hat{\Kbar}^\dagger\tau}\right].
 \label{S:eq:yieldprob}
\end{equation}
Thus two realizations differing only in uniform shift have identical normalized conditional dynamics and the exact yield ratio
\begin{equation}
 \frac{P_{\rm nj}^{(\Gamma_1)}(\tau)}{P_{\rm nj}^{(\Gamma_2)}(\tau)}
 =e^{n(\Gamma_2-\Gamma_1)\tau}.
 \label{S:eq:yieldratio}
\end{equation}
The ratio assumes an ideal no-emission record and does not by itself determine either absolute probability. As a real-phase many-particle reference, we propagate a three-cell open chain with $t=0.5$, $J=\gamma=1$. Its finite thresholds are
\begin{equation}
 \Gamma_3^{(3)}=2.161038289024,\qquad
 \Gamma_2^{(3)}=2.479920828373.
 \label{S:eq:yieldfinitevalues}
\end{equation}
For particle number $n$, the initially occupied orbitals are $B_{1,j}$ in the last $n$ cells. Direct fixed-number fermionic propagation, including antisymmetric signs, gives at $J\tau=1.5$
\begin{center}
\begin{tabular}{c c c c}
\toprule
$n$ & $P_{\rm nj}^{(\Gamma_3)}$ & $P_{\rm nj}^{(\Gamma_2)}$ & mean trials per accepted record\\
\midrule
1 & $0.143406$ & $0.088886$ & $7.0$ versus $11.3$\\
2 & $0.029671$ & $0.011399$ & $33.7$ versus $87.7$\\
3 & $0.006057$ & $0.001442$ & $165.1$ versus $693.4$\\
\bottomrule
\end{tabular}
\end{center}
The computed ratios agree with Eq.~\eqref{S:eq:yieldratio} to relative error below $3\times10^{-13}$.

\section{Biorthogonal fixed-occupancy clock sector}
\label{S:sec:clock}

The root spectrum follows the SSH construction of Ref.~\cite{McCann2026}. The present ordinary-fermion realization differs from the dissipative Fock-parafermion model of Mastiukova et al.~\cite{Mastiukova2026}, whose jump operators are generally multiparticle Fock-parafermion operators. Here the physical loss operators are linear in ordinary fermions, and the clock interpretation below is an algebraic encoding in a conditional fixed-occupancy sector. For the phase-extended cubic model, Eq.~\eqref{S:eq:phase_rotation} preserves the corresponding encoded subspaces up to an onsite unitary and rotates their centered eigenvalues. It does not establish an equivalence of the full many-body Hilbert spaces or unconditional Liouvillians.

Away from exceptional points, let $R$ be the matrix of right one-particle eigenvectors of $\Kbar$ and $R^{-1}$ the dual left transformation. Define
\begin{equation}
 f_\alpha^{\ddagger}=\sum_a R_{a\alpha}c_a^\dagger,
 \qquad
 \widetilde f_\alpha=\sum_a(R^{-1})_{\alpha a}c_a.
 \label{S:eq:biomodes}
\end{equation}
The symbol $\ddagger$ emphasizes that $f_\alpha^{\ddagger}$ need not be the Hermitian adjoint of $\widetilde f_\alpha$. Nevertheless,
\begin{equation}
 \{\widetilde f_\alpha,f_\beta^{\ddagger}\}=\delta_{\alpha\beta},
 \qquad
 \{f_\alpha^{\ddagger},f_\beta^{\ddagger}\}
 =\{\widetilde f_\alpha,\widetilde f_\beta\}=0.
 \label{S:eq:CAR}
\end{equation}
The no-jump Hamiltonian is
\begin{equation}
 \hat K=\sum_\alpha\left(\bar\varepsilon_\alpha-\frac{i\Gamma}{2}\right)
 f_\alpha^{\ddagger}\widetilde f_\alpha.
 \label{S:eq:diagK}
\end{equation}

For an open $p$-root SSH chain, the modes organize into $L$ multiplets
\begin{equation}
 \bar\varepsilon_{j,s}=\omega^s r_j,
 \qquad
 r_j=\left(\gamma^{p-2}e_j^2\right)^{1/p},
 \quad s=0,\ldots,p-1,
 \label{S:eq:multiplets}
\end{equation}
where $e_j>0$ are the positive one-particle energies of the SSH parent~\cite{McCann2026}. Let
\begin{equation}
 n_{j,s}=f_{j,s}^{\ddagger}\widetilde f_{j,s}.
 \label{S:eq:n}
\end{equation}
These commuting idempotents have eigenvalues zero and one on the biorthogonal Fock basis.

Define the invariant right-vector subspace
\begin{equation}
 \Venc=\operatorname{span}\left\{
 |\bm s\rangle_R=\prod_{j=1}^{L}f_{j,s_j}^{\ddagger}|0\rangle:
 s_j\in\{0,\ldots,p-1\}\right\}.
 \label{S:eq:Venc}
\end{equation}
It is characterized by exactly one occupied mode per multiplet,
\begin{equation}
 \sum_{s=0}^{p-1}n_{j,s}=1\qquad(j=1,\ldots,L),
 \label{S:eq:oneper}
\end{equation}
and has dimension $p^L$. Let $P_{\rm enc}$ denote the associated biorthogonal spectral idempotent. On this subspace define
\begin{align}
 Z_j&=\sum_{s=0}^{p-1}\omega^s n_{j,s},\nonumber\\
 X_j&=\sum_{s=0}^{p-1}f_{j,s+1}^{\ddagger}\widetilde f_{j,s},
 \qquad \text{with }s+1\text{ understood modulo }p.
 \label{S:eq:clockops}
\end{align}
Their action is
\begin{equation}
 Z_j|s_j\rangle_R=\omega^{s_j}|s_j\rangle_R,
 \qquad
 X_j|s_j\rangle_R=|s_j+1\rangle_R,
 \label{S:eq:clockaction}
\end{equation}
so within $\Venc$
\begin{equation}
 Z_jX_j=\omega X_jZ_j,
 \qquad Z_j^p=X_j^p=1.
 \label{S:eq:clockalgebra}
\end{equation}
Because every state in $\Venc$ has total particle number $L$, Eq.~\eqref{S:eq:diagK} restricts to
\begin{equation}
 \left.\hat K\right|_{\Venc}
 =-\frac{i\Gamma L}{2}\one_{\rm enc}
 +\sum_{j=1}^{L}r_jZ_j.
 \label{S:eq:encoding}
\end{equation}
The second term is the Baxter-Fendley free-parafermion normal form~\cite{Baxter1989,Fendley2014,Alcaraz2021}. Equations~\eqref{S:eq:clockops}-\eqref{S:eq:encoding} specialize the established fixed-occupancy fermionic representation of clock variables~\cite{Traverso2023} to the biorthogonal eigenmode multiplets of the root chain. The map is a similarity transformation and need not be unitary. The modes indexed by $j$ are generally spatially delocalized. The encoded clock variables do not acquire parafermionic exchange statistics or braiding~\cite{Cobanera2014}, distinguishing this representation from Fock-parafermion tight-binding models~\cite{McCannFock2026}.

Conditioning on no quantum jump preserves $\Venc$ and gives
\begin{equation}
 P_{\rm enc}e^{-i\hat Kt}P_{\rm enc}
 =e^{-\Gamma Lt/2}
 \exp\left[-it\sum_jr_jZ_j\right]P_{\rm enc}.
 \label{S:eq:conditional}
\end{equation}
The corresponding $L$-particle Green function has simple poles at the free-parafermion energies. With
$C_{\bm s}^{\ddagger}=f_{1,s_1}^{\ddagger}\cdots f_{L,s_L}^{\ddagger}$ and
$\widetilde C_{\bm s}=\widetilde f_{L,s_L}\cdots\widetilde f_{1,s_1}$ in reverse annihilator order,
\begin{equation}
 \langle0|\widetilde C_{\bm s}
 (z-\hat K)^{-1}C_{\bm s}^{\ddagger}|0\rangle
 =\frac{1}{z+i\Gamma L/2-\sum_j\omega^{s_j}r_j}.
 \label{S:eq:Green}
\end{equation}
The full ordinary-fermion Fock space is larger than $\Venc$, and recycling transfers encoded density operators to lower-particle sectors. The representation is therefore restricted to conditional no-jump dynamics and identifies a spectral subset of the full loss-only Liouvillian. It does not equate the full ordinary-fermion and parafermion Hilbert spaces. Conditioning preserves an initially prepared encoded sector but does not prepare or autonomously stabilize it.

\section{Exceptional Jordan block in the decaying density sector}
\label{S:sec:exceptional}
\label{S:sec:Jordan}

\subsection{Jordan blocks in the one-particle density sector}

Suppose the one-particle no-jump matrix contains a Jordan block
\begin{equation}
 K|v_0\rangle=\varepsilon_0|v_0\rangle,
 \qquad
 K|v_r\rangle=\varepsilon_0|v_r\rangle+|v_{r-1}\rangle,
 \quad r=1,\ldots,\ell-1.
 \label{S:eq:Jchain}
\end{equation}
In the one-particle density block $\mathcal B_{1,1}$, vectorization gives
\begin{equation}
 \cL_{1,1}=-i(K\otimes\one-\one\otimes K^*).
 \label{S:eq:L11}
\end{equation}
The nilpotent part associated with a pair of size-$\ell$ blocks is
\begin{equation}
 \mathcal N=-i(N\otimes\one-\one\otimes N^*),
 \qquad N^\ell=0,
 \quad N^{\ell-1}\neq0.
 \label{S:eq:nilpotent}
\end{equation}
The two terms commute. In every monomial of $\mathcal N^{2\ell-1}$, at least one factor appears with power $\ell$, so
\begin{equation}
 \mathcal N^{2\ell-1}=0.
 \label{S:eq:nilupper}
\end{equation}
At power $2\ell-2$, the only term capable of surviving has power $\ell-1$ in both factors:
\begin{equation}
 \mathcal N^{2\ell-2}
 =(-i)^{2\ell-2}(-1)^{\ell-1}
 \binom{2\ell-2}{\ell-1}
 N^{\ell-1}\otimes(N^*)^{\ell-1}\neq0.
 \label{S:eq:nillower}
\end{equation}
The nilpotency index is therefore $2\ell-1$, proving that the largest Jordan block in this paired $\ell\times\ell$ density sector has size
\begin{equation}
 \ell_{\cL}=2\ell-1.
 \label{S:eq:order}
\end{equation}
More generally, a pair of Hamiltonian blocks of sizes $\ell_a$ and $\ell_b$ generates Liouvillian block sizes
\begin{equation}
 \ell_a+\ell_b-1,\ \ell_a+\ell_b-3,\ldots,
 |\ell_a-\ell_b|+1,
 \label{S:eq:allblocks}
\end{equation}
consistent with the established Hamiltonian-to-Liouvillian and composite-system Jordan enhancement~\cite{Wiersig2020,WiersigChen2025,Shiralieva2026}.

Analytically characterized higher-order Liouvillian exceptional points have also been obtained in quadratic dissipative fermion models. Many-body quantum-jump perturbations then open finite Liouvillian gaps with perturbation-dependent fractional-power scalings~\cite{XuYi2026}. Here the density-sector Jordan order instead follows from a prescribed no-jump block with a support-optimal passive realization.

\subsection{Persistence of the decaying exceptional point under recycling}

In the subspace $\mathcal B_{1,1}\oplus\mathcal B_{0,0}$, the full Liouvillian has the block form
\begin{equation}
 \cL=
 \begin{pmatrix}
 \cL_{1,1}&0\\
 \mathcal J_{10}&0
 \end{pmatrix}.
 \label{S:eq:blockL}
\end{equation}
For the root edge exceptional point, the eigenvalue of $\cL_{1,1}$ is $-\Gamma<0$, while the vacuum eigenvalue is zero. Since the spectra of the diagonal blocks are disjoint, the off-diagonal block can be eliminated by a block-triangular similarity transformation, equivalently by solving the associated Sylvester equation. Hence the Jordan structure at $-\Gamma$ is exactly that of $\cL_{1,1}$ within this invariant density-plus-vacuum sector. Quantum jumps populate the vacuum component of the generalized eigenoperators but do not remove the decaying block. The full Liouvillian therefore has a Jordan chain of length $2\ell-1$, so its largest Jordan block at that eigenvalue is at least this large. The chain can extend into resonant higher-particle sectors. An invariant-sector proof alone does not fix the full-space block size. The relation is specific to this loss-only sector. Hamiltonian and Liouvillian exceptional points can differ in general open systems~\cite{Minganti2019,Minganti2020,Shiralieva2026}. Recent optomechanical work makes the operational distinction explicit: its Hamiltonian exceptional point belongs to conditional no-jump evolution, whereas its Liouvillian exceptional point belongs to unconditional Lindblad dynamics~\cite{GhoshBhattacharya2026}.

\subsection{Physical survival law}

The probability to remain in the one-particle sector is
\begin{equation}
 P_1(t)=\Tr[\Pi_1\rho(t)]
 =\|e^{-iKt}|\psi_0\rangle\|^2,
 \label{S:eq:P1general}
\end{equation}
because every recycling event transfers the particle to the vacuum. If the initial state has a nonzero component along the highest generalized eigenvector $|v_{\ell-1}\rangle$, Eq.~\eqref{S:eq:Jchain} gives
\begin{equation}
 e^{-iKt}|v_{\ell-1}\rangle
 =e^{-i\varepsilon_0t}
 \sum_{q=0}^{\ell-1}\frac{(-it)^q}{q!}
 |v_{\ell-1-q}\rangle.
 \label{S:eq:Jordanprop}
\end{equation}
For $\varepsilon_0=-i\Gamma/2$, the leading survival envelope is
\begin{equation}
 P_1(t)\sim t^{2\ell-2}e^{-\Gamma t}.
 \label{S:eq:generalsurvival}
\end{equation}
It is parity even, uses a positive density matrix, and defines a particle-number survival observable.

For the cubic-root SSH chain at $t=0$, the right edge is the two-orbital block
\begin{equation}
 \Kbar_{\rm edge}=
 \begin{pmatrix}0&0\\\gamma&0\end{pmatrix}
 \label{S:eq:edgeblock}
\end{equation}
in the basis $(B_{1L},B_{2L})$. Define the shifted physical block $K_{\rm edge}=\Kbar_{\rm edge}-i\Gamma\one_2/2$. Starting from $|B_{1L}\rangle$,
\begin{equation}
 e^{-iK_{\rm edge}t}|B_{1L}\rangle
 =e^{-\Gamma t/2}
 \left(|B_{1L}\rangle-i\gamma t|B_{2L}\rangle\right),
 \label{S:eq:edgeprop}
\end{equation}
and therefore
\begin{equation}
 P_1(t)=e^{-\Gamma t}\left(1+\gamma^2t^2\right).
 \label{S:eq:edgeP}
\end{equation}
The Hamiltonian block has order two and the one-particle density-sector block has order three.

A residual reverse hopping $\delta$ changes the centered edge matrix to
\begin{equation}
 \Kbar_{\rm edge}(\delta)=
 \begin{pmatrix}0&\delta\\\gamma&0\end{pmatrix}.
 \label{S:eq:reverse}
\end{equation}
For real positive $\delta$, $\Omega=\sqrt{\gamma\delta}$ and the scaled survival is
\begin{equation}
 e^{\Gamma t}P_1(t)
 =\cos^2(\Omega t)+\frac{\gamma}{\delta}\sin^2(\Omega t).
 \label{S:eq:reverseP}
\end{equation}
The limit $\delta\to0$ is $1+\gamma^2t^2$. The exceptional polynomial is therefore robust on the finite-time window
\begin{equation}
 \Omega t\ll1
 \quad\Longleftrightarrow\quad
 t\ll(\gamma|\delta|)^{-1/2}.
 \label{S:eq:window}
\end{equation}

\subsection{Exact finite-system Liouvillian spectrum}

\begin{figure}[t]
 \centering
 \includegraphics[width=\textwidth]{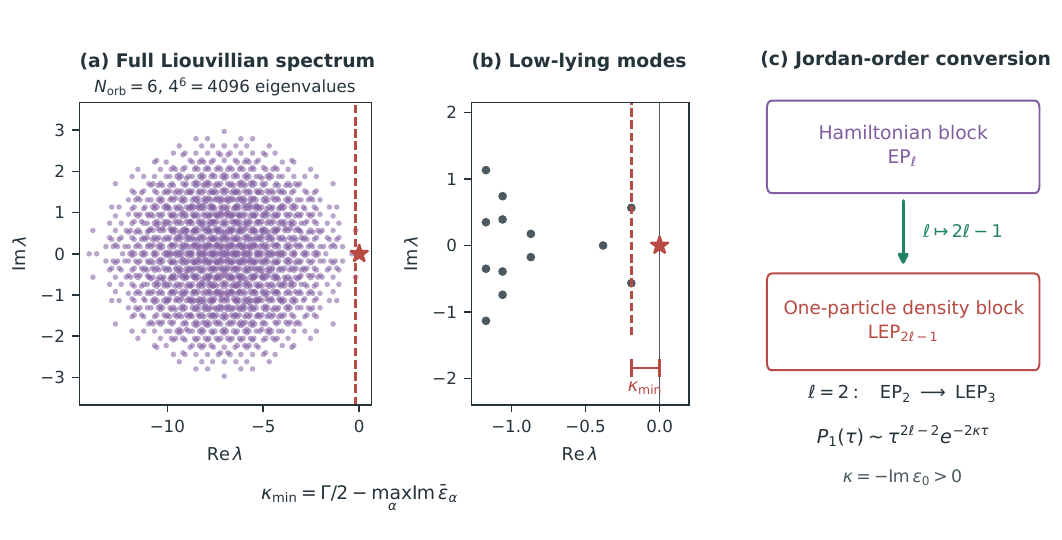}
 \caption{\label{S:fig:spectrum}
 Exact finite-system spectrum and exceptional-point block. (a) Complete $4^6=4096$ Liouvillian spectrum for the two-cell cubic-root SSH chain with the parameters in Eq.~\eqref{S:eq:S2params}. Coincident eigenvalues are combined and marker size increases logarithmically with multiplicity. The red star is the vacuum steady state and the dashed line marks the full-space gap. (b) Low-lying zoom showing $\kappa_{\min}$. The physical parity-even gap is $2\kappa_{\min}$. (c) A Hamiltonian Jordan block of order $\ell$ produces a largest one-particle density-sector block of order $2\ell-1$ and a leading survival term $P_1(t)\sim t^{2\ell-2}e^{-2\kappa t}$ for an initial state overlapping the highest generalized vector, with $\kappa=-\operatorname{Im}\varepsilon_0$. For $\ell=2$ this is an $\mathrm{LEP}_3$ block.}
\end{figure}

For the $p=3$ open chain with $L=2$ cells there are $N_{\rm orb}=6$ one-particle modes, a Fock-space dimension $2^6=64$, and an operator-space dimension $4^6=4096$. At
\begin{equation}
 t=0.5,\qquad J=\gamma=1,\qquad
 \Gamma=\Gamma_{\rm CP}=\sqrt{2(t+J)^2+\gamma^2},
 \label{S:eq:S2params}
\end{equation}
the one-particle eigenvalues of the physical no-jump matrix generate all ordinary-fermion many-body energies by subset sums. Equation~\eqref{S:eq:lambda} then gives the complete Liouvillian multiset shown in Fig.~\ref{S:fig:spectrum}. Marker size increases logarithmically with algebraic multiplicity. The vacuum is the steady-state eigenvalue at the origin, while the closest nonzero real part gives the full-operator-space gap $\kappa_{\min}$. The parity-even physical gap is twice as large. The right panel summarizes the density-sector Jordan enhancement in Eqs.~\eqref{S:eq:nilupper}-\eqref{S:eq:order}.

\section{Lossy-auxiliary implementation and robustness}
\label{S:sec:implementation}

\subsection{Adiabatic elimination}

Each collective jump can be generated by a strongly damped auxiliary mode. Let $a_\mu$ be an auxiliary fermionic orbital with damping rate $\kappa_\mu$ and coherent coupling
\begin{equation}
 H_{\rm int}=a_\mu^\dagger G_\mu+G_\mu^\dagger a_\mu,
 \qquad G_\mu=\sum_a g_{\mu a}c_a.
 \label{S:eq:Hint}
\end{equation}
We use $\Diss[L]\rho=L\rho L^\dagger-\tfrac12\{L^\dagger L,\rho\}$. The joint dynamics contains $\kappa_\mu\Diss[a_\mu]$. In the regime
\begin{equation}
 \kappa_\mu\gg \|G_\mu\|,\ \|H_{\rm sys}\|,
 \label{S:eq:adiabaticcondition}
\end{equation}
adiabatic elimination of the auxiliary vacuum gives, to leading nonvanishing order~\cite{LeRegent2024},
\begin{equation}
 \dot\rho_{\rm sys}=\frac{4}{\kappa_\mu}\Diss[G_\mu]\rho_{\rm sys}+\cdots.
 \label{S:eq:effectiveD}
\end{equation}
Thus a desired jump $L_\mu=\sum_a b_{\mu a}c_a$ is obtained by choosing
\begin{equation}
 g_{\mu a}=\frac{\sqrt{\kappa_\mu}}{2}b_{\mu a}.
 \label{S:eq:gchoice}
\end{equation}
For a translation-invariant realization the conditions become
$\sup_k\|B(k)\|^2\ll\kappa$ and
$\sup_k\|H_{\rm sys}(k)\|\ll\kappa$.
Synthetic phases in $b_{\mu a}$ can be supplied by complex coherent couplings, as in reservoir-engineered nonreciprocity and recent auxiliary-atom proposals~\cite{Metelmann2015,LiuSegal2020,Liu2025,Bai2026}. The real-phase factor uses three auxiliaries per cell. The phase-optimal factor in Eq.~\eqref{S:eq:phase_explicit_rows} uses two. The many-body clock sector requires fermionic system and auxiliary modes. The same one-particle factorization may also be emulated in bosonic photonic or circuit arrays. Phase-dependent collective dissipation also appears in driven-dissipative response and topology~\cite{Wanjura2025}. That setting addresses a different question from minimizing uniform loss under a fixed scalar-support constraint.
This static spatial factorization problem is distinct from minimizing auxiliary modes that approximate a frequency-dependent bath spectrum, for which different lower bounds apply~\cite{Xiang2026}.

\subsection{Finite auxiliary linewidth at the edge exceptional point}

A nondefective bulk benchmark does not determine the spectral scaling at an exceptional point. We therefore retain the auxiliary modes for the two-orbital edge block,
\begin{equation}
 H_{\rm e}=\frac{\gamma}{2}\sigma_x,\qquad
 M_{\rm e}=\Gamma\one_2+\gamma\sigma_y,\qquad
 K_{\rm e}=H_{\rm e}-\frac{i}{2}M_{\rm e}
 =\begin{pmatrix}-i\Gamma/2&0\\ \gamma&-i\Gamma/2\end{pmatrix}.
 \label{S:eq:edge-finite-HMK}
\end{equation}
For $\Gamma>\gamma$, a convenient square factor and the corresponding coherent auxiliary coupling are
\begin{equation}
 B=\begin{pmatrix}
 \sqrt{(\Gamma-\gamma)/2}&i\sqrt{(\Gamma-\gamma)/2}\\
 \sqrt{(\Gamma+\gamma)/2}&-i\sqrt{(\Gamma+\gamma)/2}
 \end{pmatrix},\qquad G=\frac{\sqrt\kappa}{2}B.
 \label{S:eq:edgefactor}
\end{equation}
With two system and two auxiliary modes, the exact no-emission matrix is
\begin{equation}
 K_{\rm ext}=\begin{pmatrix}H_{\rm e}&G^\dagger\\G&-i\kappa\one_2/2\end{pmatrix}.
 \label{S:eq:Kext}
\end{equation}
Its characteristic polynomial is independent of the chosen square factor:
\begin{align}
 p(z)&=\det(z\one_4-K_{\rm ext})\nonumber\\
 &=\left(z^2+\frac{i\kappa z}{2}-\frac{\kappa\Gamma}{4}\right)^2
 -\frac{\gamma^2}{4}(z^2+i\kappa z).
 \label{S:eq:edge-finite-polynomial}
\end{align}
The two slow roots $z_s$, $s=\pm1$, have the asymptotic expansion
\begin{equation}
 z_s=-\frac{i\Gamma}{2}
 -s i\gamma\sqrt{\frac{\Gamma}{2\kappa}}
 -\frac{i(\Gamma^2+\gamma^2)}{2\kappa}
 +O(\kappa^{-3/2}),
 \label{S:eq:edge-finite-roots}
\end{equation}
and therefore
\begin{equation}
 |z_+-z_-|=2\gamma\sqrt{\frac{\Gamma}{2\kappa}}
 \left[1+\frac{9\Gamma^2+\gamma^2}{4\Gamma\kappa}
 +O(\kappa^{-2})\right].
 \label{S:eq:edge-finite-splitting}
\end{equation}
Finite linewidth thus unfolds the ideal edge exceptional point at order $\kappa^{-1/2}$.

Finite-time convergence is faster and does not require diagonalizability. For initially empty auxiliaries, eliminating their amplitude gives
\begin{equation}
 \dot u(t)=-iH_{\rm e}u(t)-\frac{\kappa}{4}M_{\rm e}
 \int_0^t e^{-\kappa(t-s)/2}u(s)\,ds.
 \label{S:eq:edge-volterra}
\end{equation}
Set $m=\|M_{\rm e}\|_2$ and $C=\|H_{\rm e}\|_2+m/2$. Integration by parts, contractivity, and variation of constants give
\begin{align}
 \left\|U_{ss}(t)-e^{-iK_{\rm e}t}\right\|_2
 &\leq \frac{m}{\kappa}\left\{1-e^{-\kappa t/2}
 +C\left[t-\frac{2}{\kappa}(1-e^{-\kappa t/2})\right]\right\}\nonumber\\
 &\leq \frac{m}{\kappa}(1+Ct).
 \label{S:eq:edge-volterra-bound}
\end{align}
The auxiliary convolution also yields
\begin{equation}
 P_{\rm aux}(t)\leq\frac{m}{\kappa}(1-e^{-\kappa t/2})^2.
 \label{S:eq:edge-aux-bound}
\end{equation}

At finite linewidth, write the no-emission amplitude as $(u,a)^T=e^{-iK_{\rm ext}t}(\psi,0)^T$ and distinguish
\begin{align}
 P_{\rm sys}(t)&=\|u(t)\|^2,\nonumber\\
 P_0(t)&=P_{\rm no\,emission}(t)=\|u(t)\|^2+\|a(t)\|^2
 =P_{\rm sys}(t)+P_{\rm aux}(t).
 \label{S:eq:edge-probabilities}
\end{align}
$P_{\rm sys}$ is the primary-system number signal. The quantity $P_0$ is the probability that no auxiliary emission has occurred and includes residual auxiliary occupation. Equations~\eqref{S:eq:edge-volterra-bound} and \eqref{S:eq:edge-aux-bound} show that both converge to $P_{\rm tar}=e^{-\Gamma t}[1+(\gamma t)^2]$ as $O(\kappa^{-1})$ at fixed time.

For $\gamma=1$, $\Gamma=\sqrt3$, and $t=1/\Gamma$, independent four-mode propagation over $640\leq\kappa/\Gamma\leq10240$ gives log-log slopes $-0.50119$ for the slow-root separation, $-1.00037$ for the system-block propagator error, $-1.00122$ for $|P_{\rm sys}-P_{\rm tar}|$, and $-1.00124$ for $|P_0-P_{\rm tar}|$. At $\kappa/\Gamma=80$ and the later time $t=4/\gamma$, exact propagation gives $P_{\rm sys}=0.0162724$, $P_{\rm aux}=0.0001506$, and $P_0=0.0164230$, whereas $P_{\rm tar}=0.0166559$.

\begin{figure}[t]
 \centering
 \includegraphics[width=\textwidth]{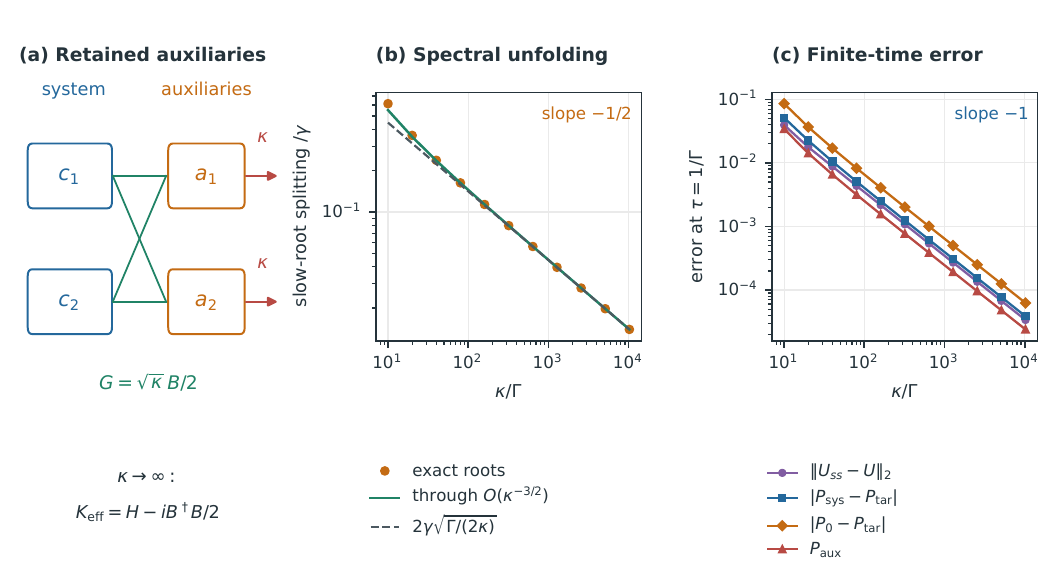}
 \caption{\label{S:fig:bandwidth}
 Finite auxiliary linewidth at the exceptional edge. (a) Two rows of $B$ couple two system modes to two lossy auxiliaries. (b) The exact slow-root splitting follows Eq.~\eqref{S:eq:edge-finite-splitting}. Finite $\kappa$ lifts the target degeneracy as $\kappa^{-1/2}$. (c) At $t=1/\Gamma$, the system-block propagator error, the two operational survival errors, and the residual auxiliary occupation all converge as $\kappa^{-1}$. The curves use the exact four-mode matrix rather than the adiabatically eliminated generator.}
\end{figure}

\subsection{Implementation errors}

Let the implemented coherent and spectral factors be $H+\delta H$ and $B+\delta B$, with $\delta H=\delta H^\dagger$. Complete positivity is preserved exactly, because the implemented loss matrix remains $(B+\delta B)^\dagger(B+\delta B)\succeq0$. The no-jump error is
\begin{equation}
 \delta K=\delta H-\frac{i}{2}
 \left(B^\dagger\delta B+\delta B^\dagger B+\delta B^\dagger\delta B\right),
 \label{S:eq:deltaK}
\end{equation}
with the norm bound
\begin{equation}
 \|\delta K\|\leq\|\delta H\|+\|B\|\,\|\delta B\|
 +\frac12\|\delta B\|^2.
 \label{S:eq:errorbound}
\end{equation}
For the affine symmetry, a convenient dimensionless diagnostic is
\begin{equation}
 \mathcal E_{\rm aff}=
 \frac{\|Z(K+i\Gamma\one/2)Z^{-1}
 -\omega(K+i\Gamma\one/2)\|_F}
 {\|K+i\Gamma\one/2\|_F}.
 \label{S:eq:afferror}
\end{equation}
Equation~\eqref{S:eq:errorbound} implies linear sensitivity to small coefficient errors, while Eq.~\eqref{S:eq:window} gives the independent time-domain tolerance of the exceptional transient. Finite auxiliary linewidth produces corrections controlled by the small ratios in Eq.~\eqref{S:eq:adiabaticcondition}.

\subsection{Contraction bound for measured survival}

Write $U(\tau)=e^{-iK\tau}$ and $\widetilde U(\tau)=e^{-i(K+\delta K)\tau}$. For the coefficient errors above, both generators are passive, so $\|U(\tau)\|_2,\|\widetilde U(\tau)\|_2\leq1$ for $\tau\geq0$. Duhamel's identity gives
\begin{equation}
 \widetilde U(\tau)-U(\tau)
 =-i\int_0^\tau\widetilde U(\tau-s)\delta K U(s)\,ds,
 \qquad
 \|\widetilde U(\tau)-U(\tau)\|_2\leq\tau\|\delta K\|_2.
 \label{S:eq:duhamel}
\end{equation}
For every normalized pure one-particle initial state, $P_1=\|U\psi\|^2$. Factoring the difference of these quadratic forms and using both contraction bounds yields
\begin{equation}
 |\widetilde P_1(\tau)-P_1(\tau)|
 \leq\min\{1,2\tau\|\delta K\|_2\}.
 \label{S:eq:survivalerror}
\end{equation}
Convexity extends this to mixed initial states. Combining with Eq.~\eqref{S:eq:errorbound} provides a bound in terms of the implemented Hamiltonian and jump coefficients. It holds without diagonalizability, a spectral gap, or a bound on eigenvector condition numbers. Consequently square-root exceptional eigenvalue splitting coexists with Lipschitz finite-time survival. The bound controls absolute error. Exponentially rescaling a small survival signal can still magnify relative errors at late times.

\subsection{Connected-chain coefficient-error ensemble}

\begin{figure}[t]
 \centering
 \includegraphics[width=0.98\textwidth]{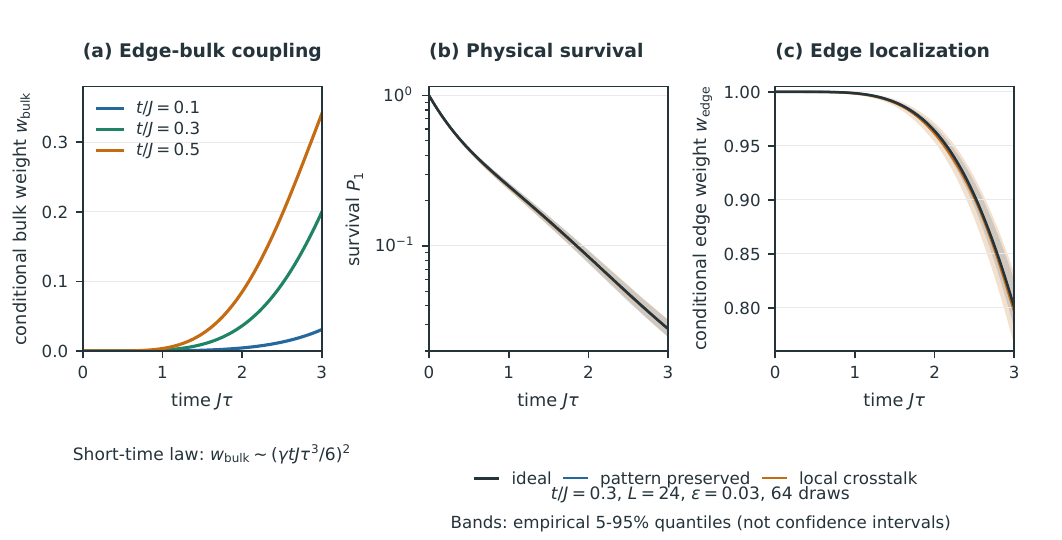}
 \caption{\label{S:fig:robustness}
 Connected-chain robustness. (a) Conditional bulk leakage for $t/J=0.1,0.3,0.5$ and $L=24$, proving that the initialized edge is dynamically coupled to the bulk. (b) Absolute survival for $t/J=0.3$, $L=24$, and $\epsilon=0.03$. (c) Conditional weight in the last unit cell for the same realization. Black curves are ideal. Colored curves are medians for pattern-preserving and local-crosstalk errors. Bands are empirical 5-95\% quantiles of 64 draws and are not confidence intervals.}
\end{figure}

The exact edge law at $t=0$ lives in an invariant two-mode block, so perturbing that limit does not by itself test edge-to-bulk transport. Figure~\ref{S:fig:robustness} instead uses the open cubic-root chain at $J=\gamma=1$, $t/J=0.1,0.3,0.5$, and $L=6,12,24$. Each sample is shifted by its exact finite threshold
\begin{equation}
 \Gamma_L=\sqrt{\gamma^2+2\|t\one_L+JS_L\|_2^2}
 \label{S:eq:connectedthreshold}
\end{equation}
and factorized as $M_L=B_L^\dagger B_L$ with $3L-1$ rows of cell range one. Starting from $|B_{1,L}\rangle$, the edge region is the last complete unit cell. We record
\begin{equation}
 P_1=\|\psi(\tau)\|^2,\quad
 P_{\rm edge}=\|\Pi_{\rm edge}\psi(\tau)\|^2,\quad
 w_{\rm edge}=P_{\rm edge}/P_1,\quad
 w_{\rm bulk}=1-w_{\rm edge}.
 \label{S:eq:connectedobservables}
\end{equation}
The centered generator is strictly connected: $\Pi_{\rm bulk}\Kbar^m|B_{1,L}\rangle$ vanishes for $m=1,2$ and has norm $\gamma tJ$ for $m=3$. Therefore
\begin{equation}
 w_{\rm bulk}(\tau)=\frac{\gamma^2t^2J^2}{36}\tau^6+O(\tau^7).
 \label{S:eq:connectedshorttime}
\end{equation}
This analytic coefficient and the full propagation are independently verified for every plotted size.

Two physical error ensembles are used. In the \emph{pattern-preserving} ensemble, every designed nonzero coefficient of $H$ and $B$ receives an independent relative amplitude and phase error uniformly distributed in $[-\epsilon,\epsilon]$. The \emph{local-crosstalk} ensemble adds random coefficients on previously zero $H$ and $B$ entries within the same cell range $R\leq1$, while preserving $H=H^\dagger$. Such crosstalk can populate up to six scalar coefficients in a row and therefore tests leakage outside the three-mode support class. Both implement the loss as $\widetilde B^\dagger\widetilde B$, so complete positivity is exact in every realization. For each configuration and $\epsilon=0.01,0.03$, 64 fixed-seed draws are propagated. The bands are empirical 5-95\% sample quantiles, not confidence intervals or adversarial bounds. These real-phase single-architecture checks are supplementary diagnostics. The Letter's architecture comparison is the separately matched phase-optimal calculation.

At the representative connected point $t/J=0.3$, $L=24$, $\epsilon=0.03$, and $J\tau=3$, the ideal values are $P_1=0.0280931$ and $w_{\rm edge}=0.800612$. Pattern-preserving errors give median $P_1=0.0281410$ with 5-95\% interval $[0.0247552,0.0322413]$ and median $w_{\rm edge}=0.799263$ with interval $[0.781491,0.825837]$. Local crosstalk gives $P_1=0.0282123$ with interval $[0.0248882,0.0323614]$ and $w_{\rm edge}=0.796235$ with interval $[0.765755,0.831782]$. The minimum computed loss eigenvalue over all draws is $-2.86\times10^{-15}$, consistent with roundoff, and the largest observed ratio to the Duhamel survival bound in Eq.~\eqref{S:eq:survivalerror} is $0.806$.

\subsection{Preparation, readout, and finite detector efficiency}

A one-excitation protocol can use qubits, hard-core bosons, or resonators, since particle statistics do not affect this sector. Starting from vacuum in the system and auxiliaries, prepare $|B_{1,L}\rangle$, switch on the coherent and system-auxiliary couplings, and evolve over a calibrated time grid. At each final time, switch off the couplings and perform a site-resolved binary number measurement. This readout estimates $P_{\rm sys}$ and resolves the edge and bulk weights. The auxiliary output record provides a separate calibration observable.

With output-detection efficiency $\eta$ and independent total dark-count rate $d$, at most one true emission occurs in the one-excitation protocol. The no-click probability is
\begin{equation}
 P(C_0)=e^{-d\tau}[P_0+(1-\eta)(1-P_0)]
 =e^{-d\tau}[1-\eta+\eta P_0],
 \label{S:eq:noclick}
\end{equation}
and the purity of the no-click ensemble is
\begin{equation}
 \Pr(0\ \text{emissions}\mid C_0)
 =\frac{P_0}{P_0+(1-\eta)(1-P_0)}.
 \label{S:eq:noclickpurity}
\end{equation}
This is the finite-efficiency hybrid-Liouvillian effect discussed in Ref.~\cite{Minganti2020}. At $\gamma\tau=4$ in the eliminated limit, $P_0=P_{\rm tar}=0.0166559$: Eq.~\eqref{S:eq:noclickpurity} is only $25.3\%$ for $\eta=0.95$ and $62.9\%$ for $\eta=0.99$. A purity of $95\%$ requires $\eta\geq0.999109$. No-click heralding alone is therefore unsuitable at this late time.

A final number measurement removes missed-emission contamination for one initial particle and loss only. If $F_{\rm prep}$ is the preparation fidelity, $f_1=\Pr(R=1|N_{\rm sys}=1)$, and $\epsilon_0=\Pr(R=1|N_{\rm sys}=0)$, then without conditioning on the click record
\begin{equation}
 r(\tau)=\Pr(R=1)=\epsilon_0+(f_1-\epsilon_0)F_{\rm prep}P_{\rm sys}(\tau),
 \qquad
 \widehat P_{\rm sys}=\frac{\widehat r-\epsilon_0}{(f_1-\epsilon_0)F_{\rm prep}}.
 \label{S:eq:readoutmodel}
\end{equation}
For $F_{\rm prep}=f_1=0.99$ and $\epsilon_0=10^{-3}$, an ideal binomial 95\% interval with half-width $0.1P_{\rm tar}$ at $\gamma\tau=4$ requires approximately $2.46\times10^4$ shots. A ten-time-point scan at $2.5\times10^4$ shots per point resolves the quadratic transient and a programmed reverse-hopping control $\delta/\gamma=0.05$ under the stated calibrated model. These shot estimates exclude systematic uncertainty in $\Gamma,\gamma,\kappa$, phases, and the confusion matrix. An experimental analysis would include these quantities as independently calibrated nuisance parameters in a joint binomial likelihood.

\section{Numerical and plotting details}
\label{S:sec:numerics}

\paragraph{Propagation and observables.}
The calculations use finite-dimensional matrices in the physical orbital basis and double-precision complex arithmetic. Markovian evolution is evaluated from the effective no-jump generator, whereas finite-bandwidth calculations retain the auxiliary modes explicitly and evaluate $\exp(-iK_{\rm ext}\tau)$. Initially empty auxiliaries are represented by zero amplitudes. The propagated state remains unnormalized when evaluating system survival, auxiliary population, emitted-particle probability, and target-projection yield. Normalization is applied only when forming conditional states or fidelities. The full-propagator comparison uses the system-to-system block of the extended evolution operator, with the same target matrix for both reservoir architectures.

\paragraph{Main-figure calculations.}
Figure~1 evaluates the analytic support thresholds of the \hyperref[S:sec:phase]{phase-controlled construction} over the cycle phase and $\gamma/S$. The map uses a uniform grid of 241 phase values over $-0.5\leq\phi/\pi\leq1.5$ and 161 values over $0\leq\gamma/S\leq2.5$, with the displayed vertical range restricted to $\gamma/S\leq2$. The line plot fixes $\gamma/S=1$. Each plotted reduction compares pair and collective reservoirs for the same prescribed generator at that parameter point. Figure~2 uses the \hyperref[S:sec:device]{connected six-cell protocol} at fixed $\kappa/S=1920$. Its two series comprise 64 paired link-error realizations and their extension by a common phase error and auxiliary detuning. The same collective-device realization supplies the support witness and target-yield data. Synthetic detector counts are sampled from the propagated probabilities. Exact binomial intervals, detector inversion, and the stated calibration allowances give the reported bounds. The total failure probability $0.01$ is divided among all 5120 probability groups, yielding at least $99\%$ simultaneous coverage conditional on those allowances. Empirical disorder percentiles and confidence bounds therefore describe different sources of variation.

Figure~3 uses nominal matrices without disorder or detector sampling. Its extended generator is a $46\times46$ matrix with 18 system modes and a common bank of 28 auxiliaries. For $\kappa/S=80,320,1920$, propagators are evaluated at the 30 times $S\tau=0.05,0.10,\ldots,1.50$. The figure displays the interval $S\tau\leq1$ and the bandwidth comparison at $S\tau=1$. Singular values and matrix norms determine the scalar-invariant mismatch, conditional-fidelity bounds, and target-yield enclosures derived in the \hyperref[S:sec:projective-dynamics]{full-propagator analysis}. These bounds hold for every one-particle input at each sampled time. The 1160 diagnostic inputs comprise 18 orbital states, 612 equal-weight two-mode superpositions with four relative phases, 512 Haar-distributed states, and 18 witness eigenvectors. Their sample minima are not substituted for bounds over all inputs. The separate \hyperref[S:sec:controls]{scalar-control study} reports the largest feasible yields found within its stated search family, without a claim of global optimality.

\paragraph{Supplemental spectra and dynamics.}
Figure~\ref{S:fig:bond} shows the analytic bond construction and affine spectral shift. Figure~\ref{S:fig:spectrum}(a,b) is obtained by enumerating the 64 fermionic subset sums of six one-particle rapidities and then all 4096 corresponding Liouvillian eigenvalues. Coincident eigenvalues are grouped only for display, with marker size encoding algebraic multiplicity. Gap calculations retain the full multiset. The many-particle yield calculation constructs fixed-particle-number matrices with the fermionic creation and annihilation signs and initially occupies the last $n$ orbitals of type $B_1$. Using the same centered propagator for both uniform shifts reproduces Eq.~\eqref{S:eq:yieldratio}.

The edge-bandwidth calculation propagates the retained $4\times4$ system-auxiliary block over $10\leq\kappa/\Gamma\leq10240$. Power-law fits use only $\kappa/\Gamma\geq640$. Figure~\ref{S:fig:robustness}(a) shows ideal bulk leakage for $t/J=0.1,0.3,0.5$ at $L=24$. Panels (b,c) use $t/J=0.3$, $\epsilon=0.03$, and 64 draws for each of the pattern-preserving and local-crosstalk error ensembles. Their bands are empirical 5th to 95th percentiles, not confidence intervals. Fixed pseudorandom seeds specify all stochastic ensembles, with master seeds 2609081920 for the connected-device protocol, 260909319 for the propagator diagnostic inputs, and 260906031 for the real-phase robustness calculation.

\paragraph{Numerical validation.}
Checks compare loss factors with their prescribed Gram matrices, verify the scalar support of each channel, and test preparation and readout unitarity. Eighty random phase-model instances test the analytic bulk and finite-chain thresholds, spectral rotations, and dark-mode conditions, with a largest absolute residual below $9.5\times10^{-12}$. For three orbital inputs at $S\tau=1$ and $\kappa/S=1920$, direct integration with DOP853 at relative and absolute tolerances $2\times10^{-11}$ and $2\times10^{-13}$ agrees with matrix exponentiation to $7.1\times10^{-13}$ in amplitude. The polar-unitary construction saturates the coherent mismatch bound to $2.8\times10^{-15}$. The retained edge calculation has a relative characteristic-polynomial residual below $3\times10^{-13}$. Additional checks cover parity-resolved gaps, clock algebra, Jordan indices, witness reconstruction, calibration bounds, and the probability ordering $0\leq Y\leq P_{\rm sys}\leq P_{\rm sys}+P_{\rm aux}\leq1$. These finite checks assess the numerical implementations and examples. The general bounds follow from the analytic derivations.

\paragraph{Plotting and data availability.}
All figures are vector plots of the calculated quantities or diagrams of the analytic construction. No smoothing is applied to the plotted numerical values. Lines between sampled values guide the eye and do not extend discrete-time bounds to intermediate times. Shaded data bands denote either the matrix-norm enclosures or empirical percentiles specified in the corresponding caption. All reported records are numerical calculations or synthetic counts, with no experimental measurements. The numerical data underlying the figures and the code used for the calculations and figure generation are available from the corresponding author upon reasonable request.

\bibliography{references}